%% file: main_and_appendix.tex
\documentclass[pmlr]{jmlr}



\usepackage{longtable}

\usepackage{booktabs}
\usepackage{siunitx} 

\theorembodyfont{\upshape}
\theoremheaderfont{\scshape}
\theorempostheader{:}
\theoremsep{\newline}

\usepackage{amssymb}            
\usepackage{mathtools}          
\usepackage{mathrsfs}           
\usepackage{graphicx}           
\usepackage{subcaption}         
\usepackage[space]{grffile}     
\usepackage{url}                
\usepackage{lipsum}             

\usepackage{multirow}
\usepackage{microtype}
\usepackage{comment}
\usepackage{placeins} 
\usepackage{etoolbox}

\makeatletter
\patchcmd{\enddocument}{\clearpage}{\FloatBarrier}{}{}
\patchcmd{\enddocument}{\@clearpage}{\FloatBarrier}{}{}
\makeatother

\usepackage[normalem]{ulem} 
\usepackage{xcolor}

\newif\ifhighlight
\highlightfalse  

\newcommand{\rev}[1]{%
    \ifhighlight
        \textcolor{red}{#1}%
    \else
        #1%
    \fi
}

\newcommand{\del}[1]{%
    \ifhighlight
        \textcolor{purple}{\sout{#1}}%
    \else
    \fi
}

\newcommand{\proofread}[1]{%
    \ifhighlight
        \textcolor{orange}{#1}%
    \else
        #1%
    \fi
}

\makeatletter
\renewcommand*{\@jmlrenddoc}{\label{jmlrend}}
\makeatother

\jmlrvolume{349}
\jmlryear{2026}
\jmlrworkshop{Conference on Applied Machine Learning for Information Security}

\title[Alignment for Cyber Defense]{Alignment of Autonomous Cyber Defense
Agents via Adaptive Policy Shaping}

  \author{\Name{Dena Mujtaba} \Email{dena.mujtaba@kitware.com}\\
   \Name{Bharadwaj Ravichandran} \Email{barry.ravichandran@kitware.com}
   \\
   \Name{Brian Hu} \Email{brian.hu@kitware.com} \\
   \Name{Arslan Basharat} \Email{arslan.basharat@kitware.com}
   \\
   \addr Kitware, 1712 Route 9 Suite 300, Clifton Park, NY USA}

\editor{Edward Raff and Ethan M. Rudd}

\begin{document}

\maketitle

\input{sections/abstract}


\title{Alignment of Autonomous Cyber Defense
Agents via Adaptive Policy Shaping}

\input{sections/introduction}
\input{sections/related_work}
\input{sections/methods}
\input{sections/results}
\input{sections/conclusion}

\acks{\rev{This material is based upon work supported by the Defense Advanced Research Projects Agency and the Air Force Research Laboratory, contract number(s): FA8650-23-C-7316. Any opinions, findings and conclusions, or recommendations expressed in this material are those of the author(s) and do not necessarily reflect the views of AFRL or DARPA.}}

\bibliography{main}

\newpage
\appendix
\renewcommand\thefigure{S\arabic{figure}} 
\setcounter{figure}{0}    
\renewcommand\thetable{S\arabic{table}}    
\setcounter{table}{0}    
\section{\protect\rev{CybORG} CAGE 2 Scenario Action Space}

\begin{figure}[h!]
\centering
\includegraphics[width=\columnwidth]{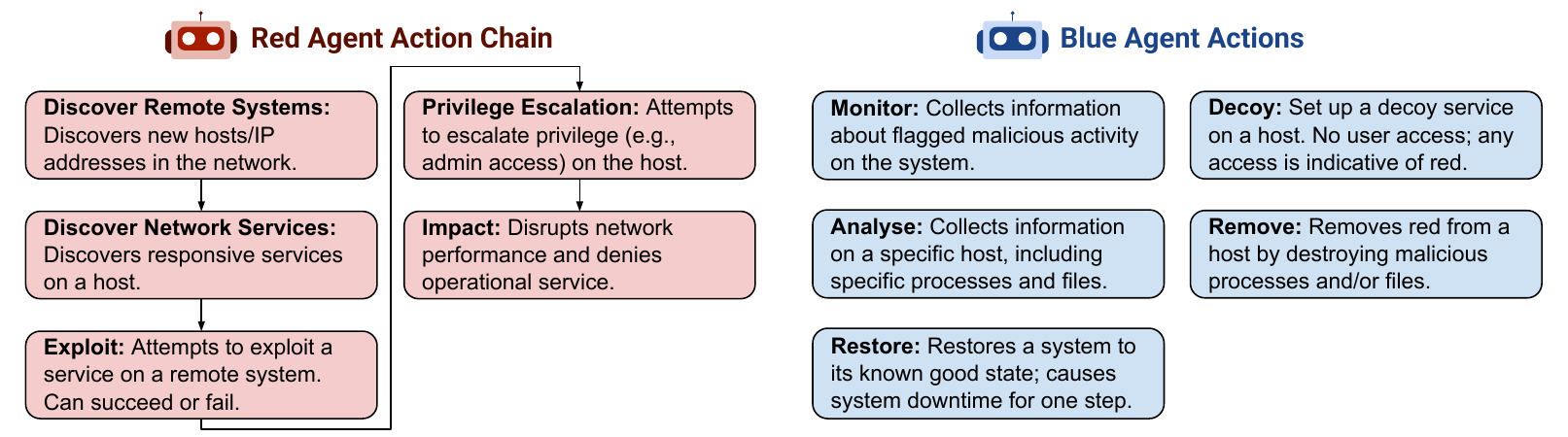}
\caption{Available red and blue agent actions in the simulated \protect\rev{CybORG} cyber defense environment.}
\label{fig:agents}
\end{figure}

Fig.~\ref{fig:agents} shows the set of defensive actions for the blue agent and the ordered list of actions along the attack chain for the red agent. 

\section{Blue Agent Performance Across Network Scales}


\rev{Tabs.~\ref{tbl:base}--\ref{tbl:hosts_4} report blue agent performance against the M-CIA, M-CAGE, and B-line red agents across four network scales.} Each table reports per-attribute alignment, reward\del{, CIA impact} and resilience drop for all alignment methods using the Meander and B-line red agents. From the values presented in these scaled scenario tables, we are able to see how \rev{training-time alignment deteriorates as network size increases}, while test-time policy shaping maintains consistent alignment across all scales.

\input{tables/base}
\input{tables/hosts_2}

\input{tables/hosts_3}

\input{tables/hosts_4}

\FloatBarrier

\section{Training-Time Reward Shaping Curves}

\begin{figure}[h!]
\centering
\includegraphics[width=\columnwidth]{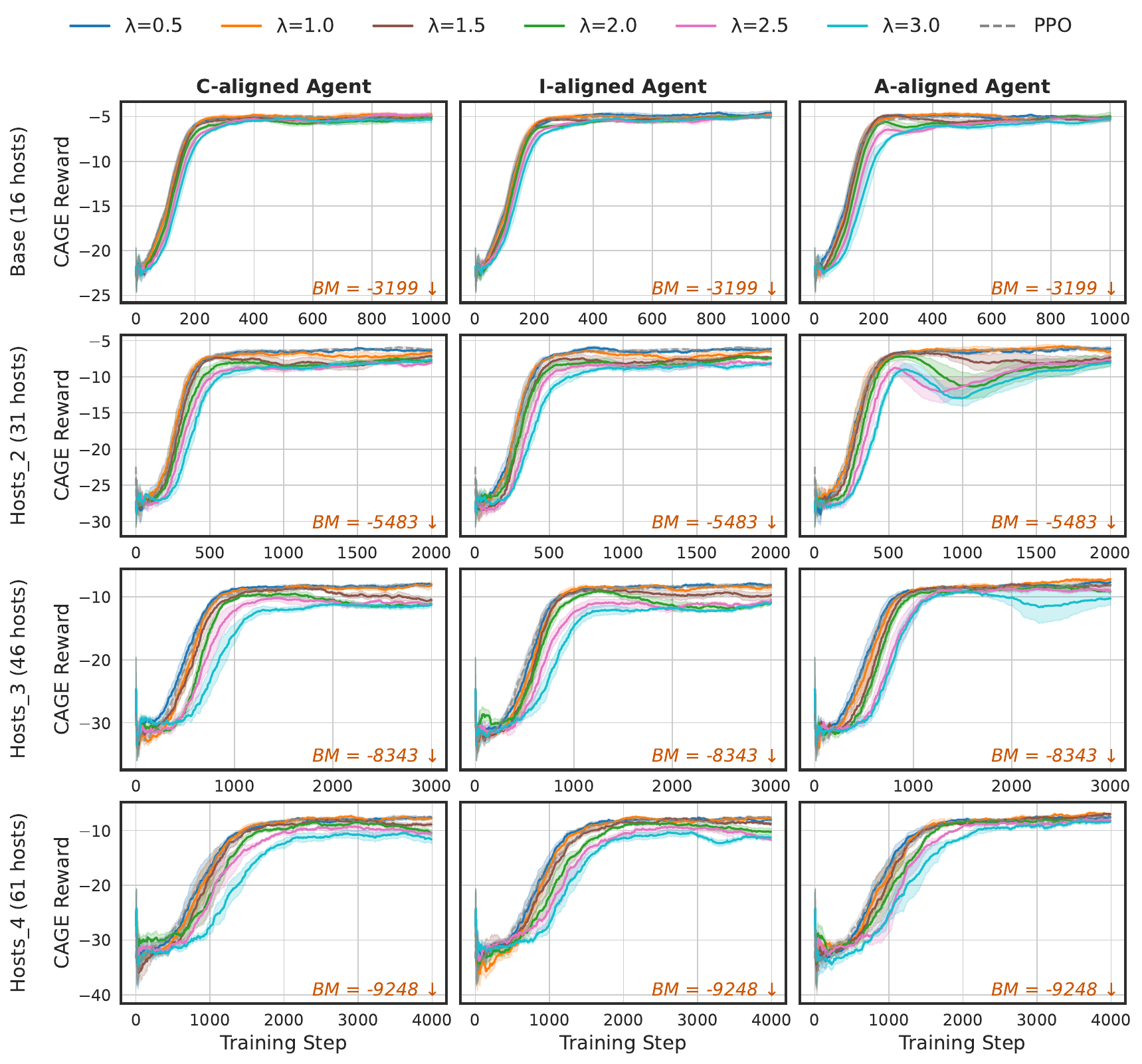}
\caption{CAGE reward during training for training-time aligned agents across network sizes (rows) and alignment attributes (columns). \del{Blue and orange}\rev{Colored} curves show agents trained with reward shaping at $\lambda \in [0.5, 3.0]$. The gray dashed line shows the PPO baseline (no alignment) for reference. The Blue Monitor baseline reward is annotated with ``BM $= x$ $\downarrow$''.}
\label{fig:RL_training_reward_CAGE_reward_subplots_full}
\end{figure}

\begin{figure}[h!]
\centering
\includegraphics[width=\columnwidth]{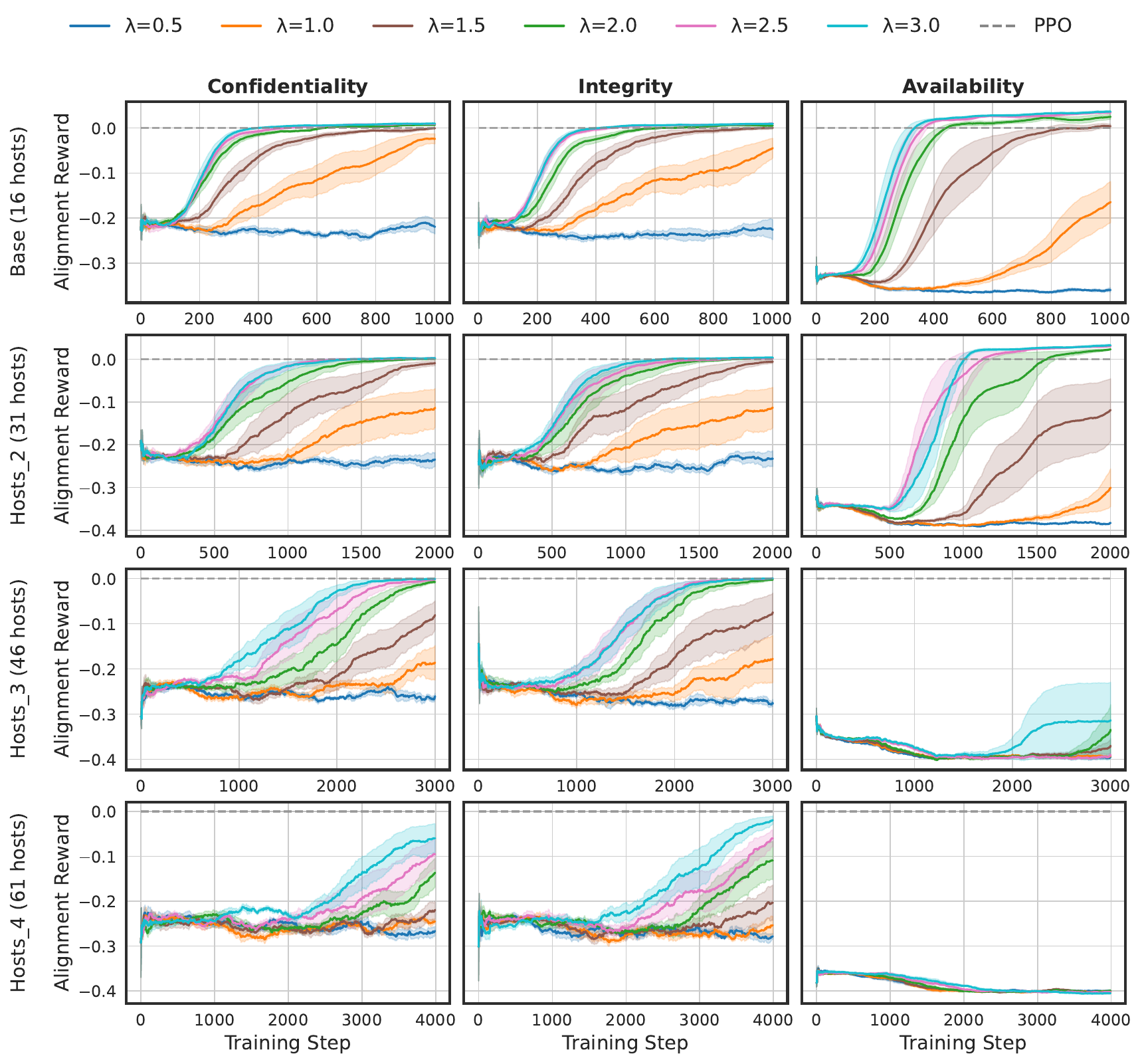}
\caption{Alignment reward during training for training-time aligned agents across network sizes (rows) and alignment attributes (columns). \del{Blue and orange}\rev{Colored} curves show agents trained with alignment reward shaping at $\lambda \in [0.5, 3.0]$. The gray dashed line shows the PPO baseline (zero alignment reward by construction).}
\label{fig:RL_training_reward_Alignment_reward_subplots}
\end{figure}

Figs.~\ref{fig:RL_training_reward_CAGE_reward_subplots_full} and~\ref{fig:RL_training_reward_Alignment_reward_subplots} show training curves for CAGE reward and alignment reward, respectively, across all four network scales, illustrating how training-time alignment reward convergence degrades with increasing network size.

\section{Blue Agent Gameplay Results}

\begin{figure}[h!]
\centering
\includegraphics[trim={0 0.8cm 0 0},clip,width=0.75\textwidth]{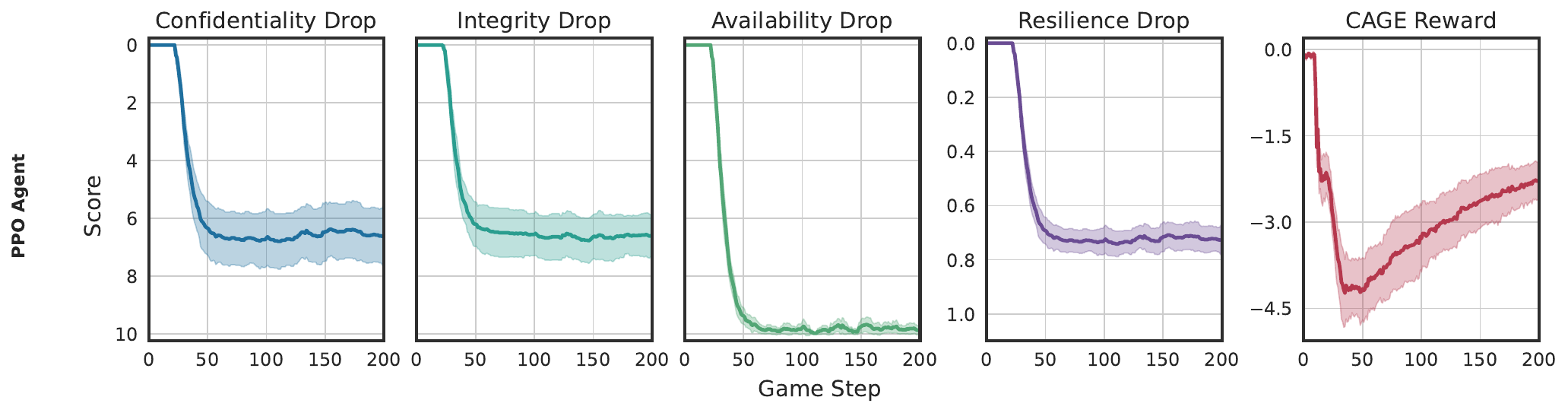}
\includegraphics[trim={0 0.8cm 0 0},clip,width=0.75\textwidth]{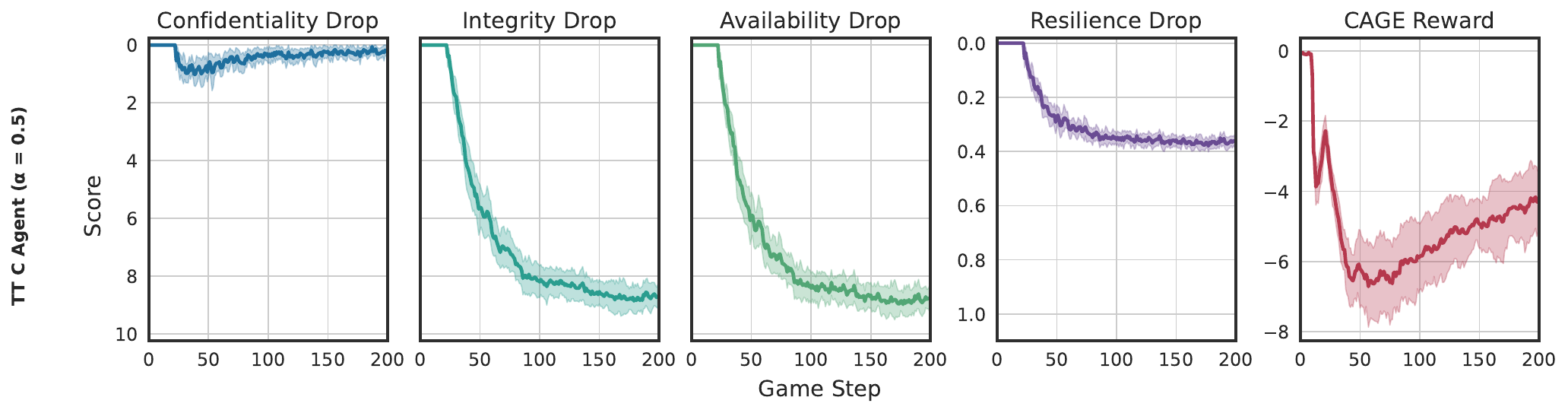}
\includegraphics[trim={0 0.8cm 0 0},clip,width=0.75\textwidth]{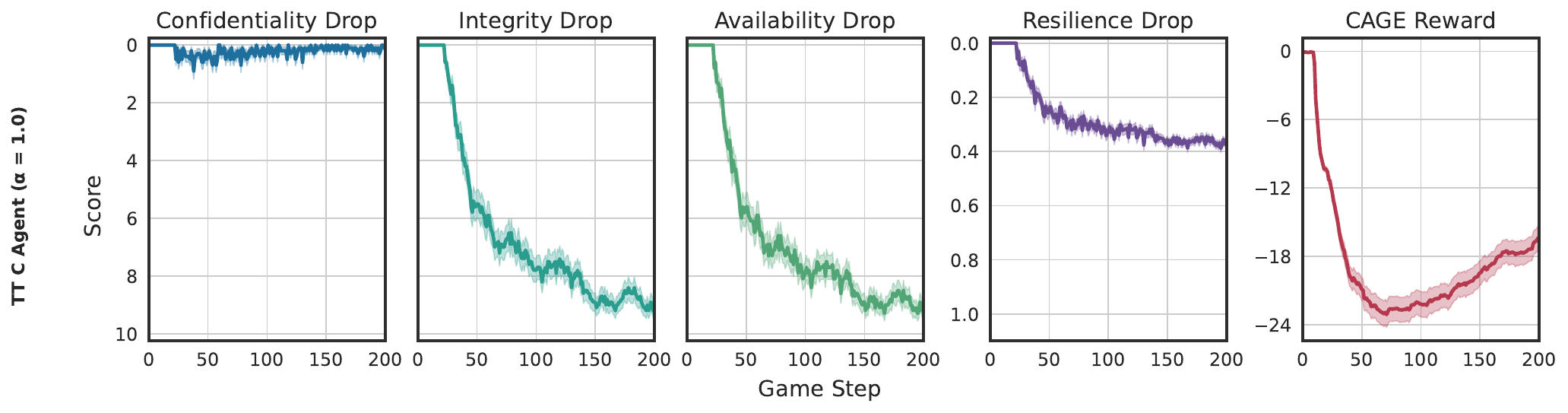}
\includegraphics[trim={0 0.8cm 0 0},clip,width=0.75\textwidth]{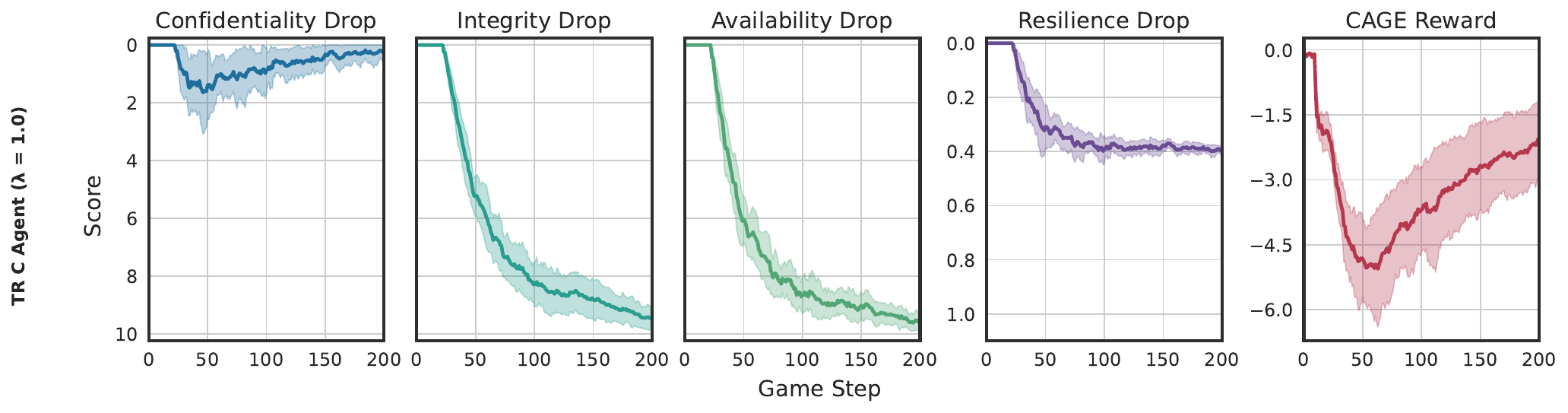}
\includegraphics[trim={0 0.8cm 0 0},clip,width=0.75\textwidth]{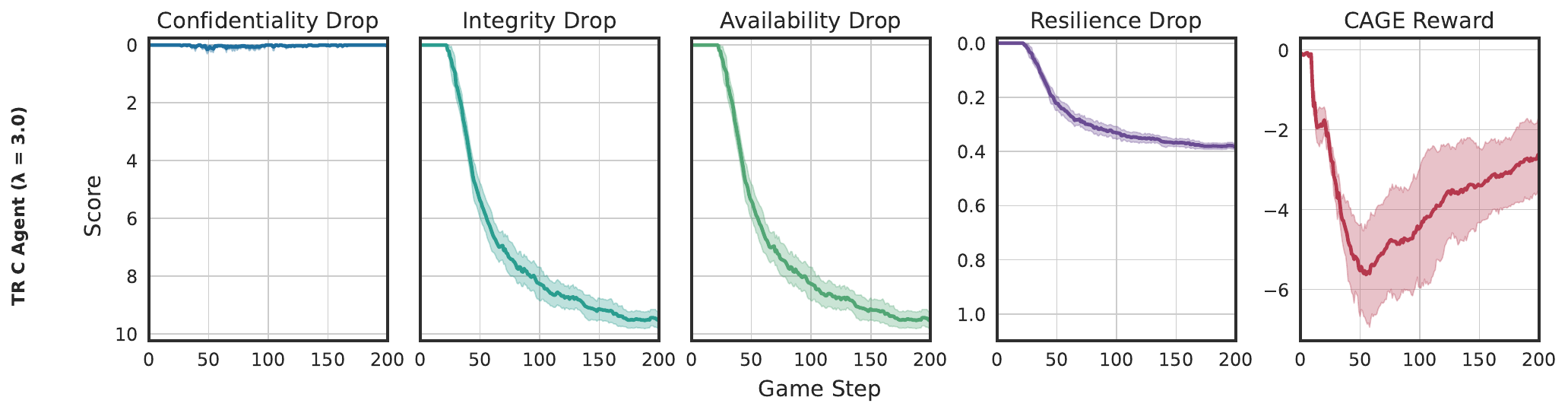}
\includegraphics[width=0.75\textwidth]{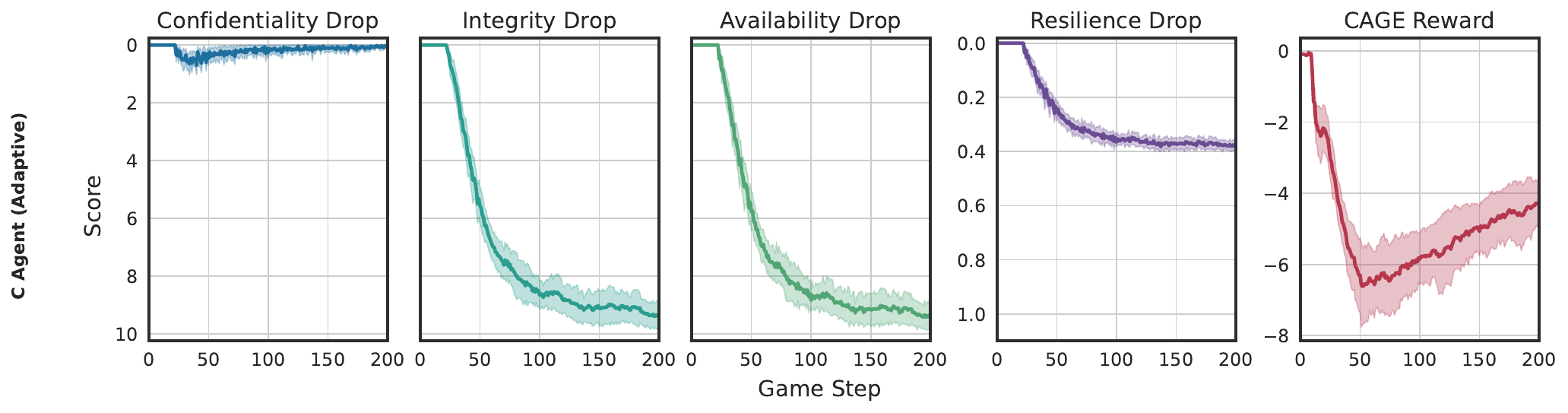}

\caption{Comparing alignment approaches for the Confidentiality agent in the Base scenario (mean $\pm$ SEM, 100 episodes\protect\rev{)}. Row 1: PPO baseline. Rows 2--3: test-time policy shaping (TT) with $\alpha=0.5$ and $\alpha=1.0$, respectively. Rows 4--5: training-time reward shaping (TR) with $\lambda=1.0$ and $\lambda=3.0$, respectively. Row 6: adaptive test-time policy shaping. At Base scale, training-time $\lambda=3.0$ achieves the strongest CIA suppression at the lowest reward cost; however, training-time shaping requires a separate training run per $\lambda$ and fails at larger network sizes, whereas test-time and adaptive shaping operate on a single frozen PPO policy and scale reliably.}
\label{fig:base_gameplay_example}
\end{figure}

\begin{figure*}[h!]
    \centering
\includegraphics[trim={0 0.8cm 0 0},clip,width=0.75\textwidth]{figures/CAMLIS/appendix/base/Adaptive/200_ppo_agent.pdf}
\includegraphics[trim={0 0.8cm 0 0},clip,width=0.75\textwidth]{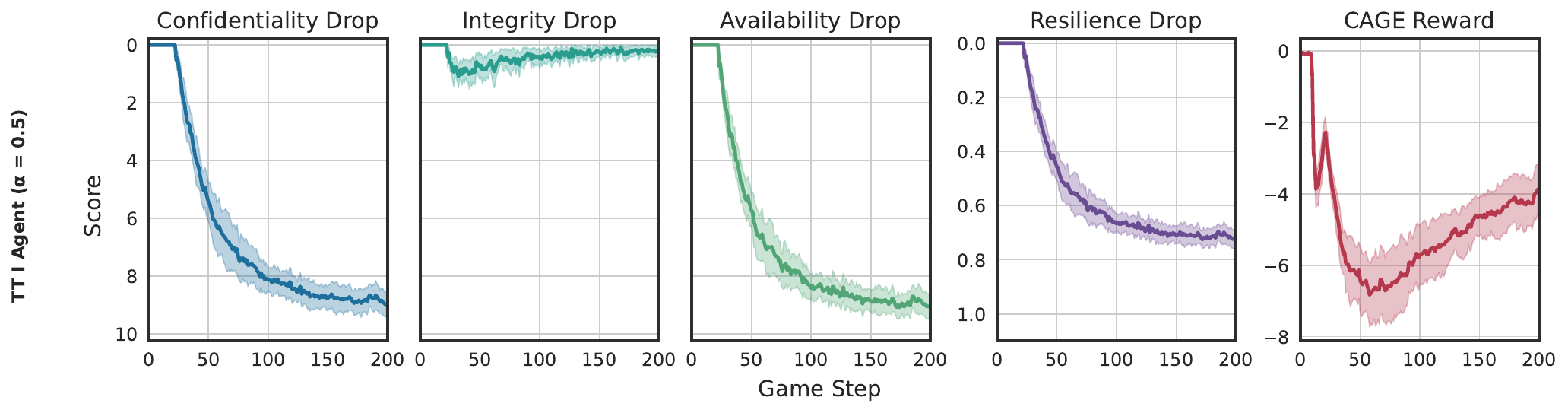}
\includegraphics[trim={0 0.8cm 0 0},clip,width=0.75\textwidth]{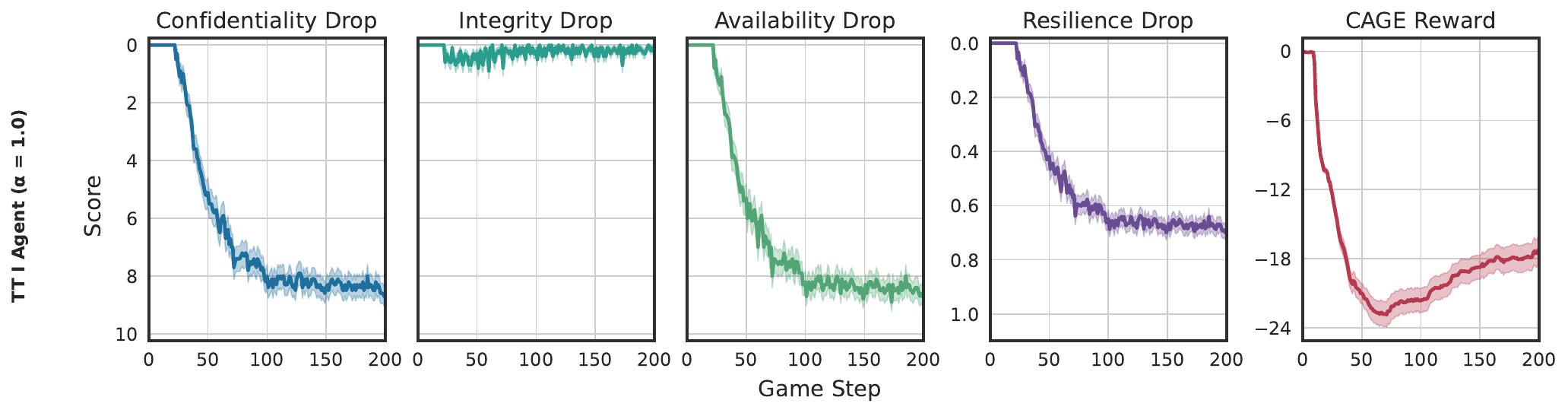}
\includegraphics[trim={0 0.8cm 0 0},clip,width=0.75\textwidth]{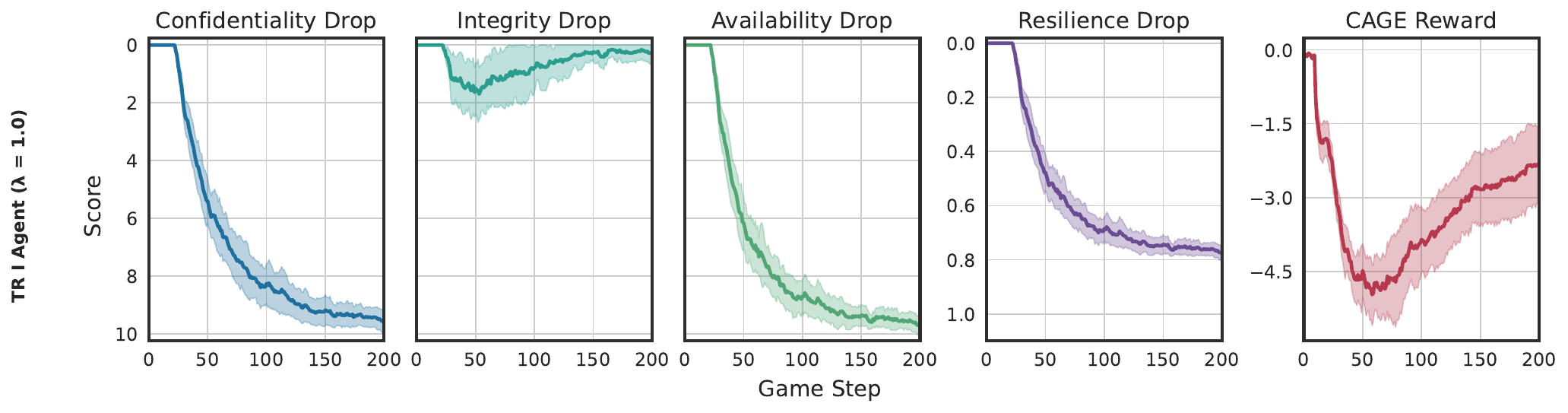}
\includegraphics[trim={0 0.8cm 0 0},clip,width=0.75\textwidth]{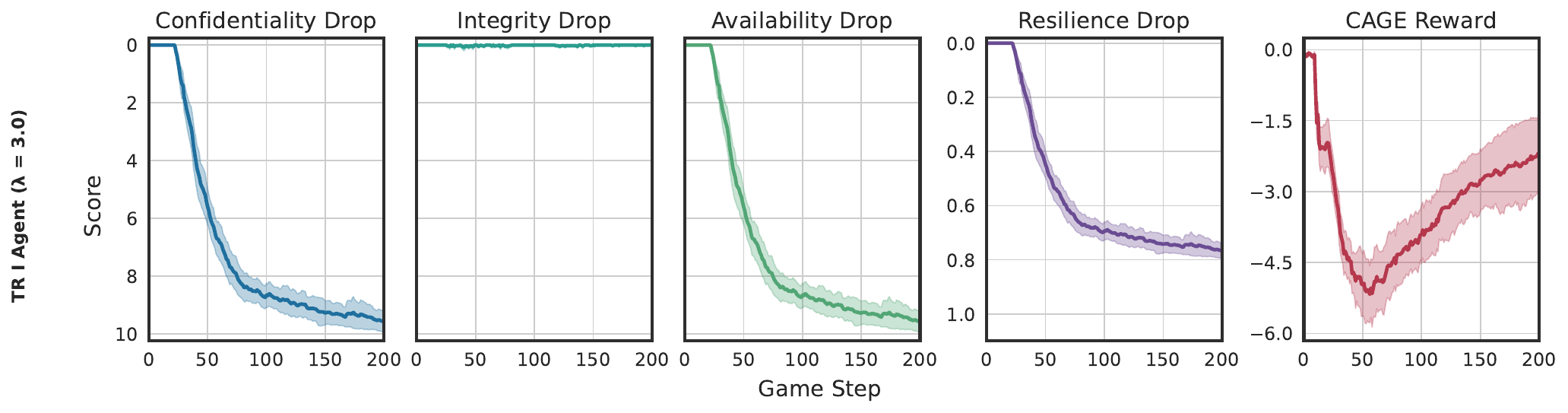}
\includegraphics[width=0.75\textwidth]{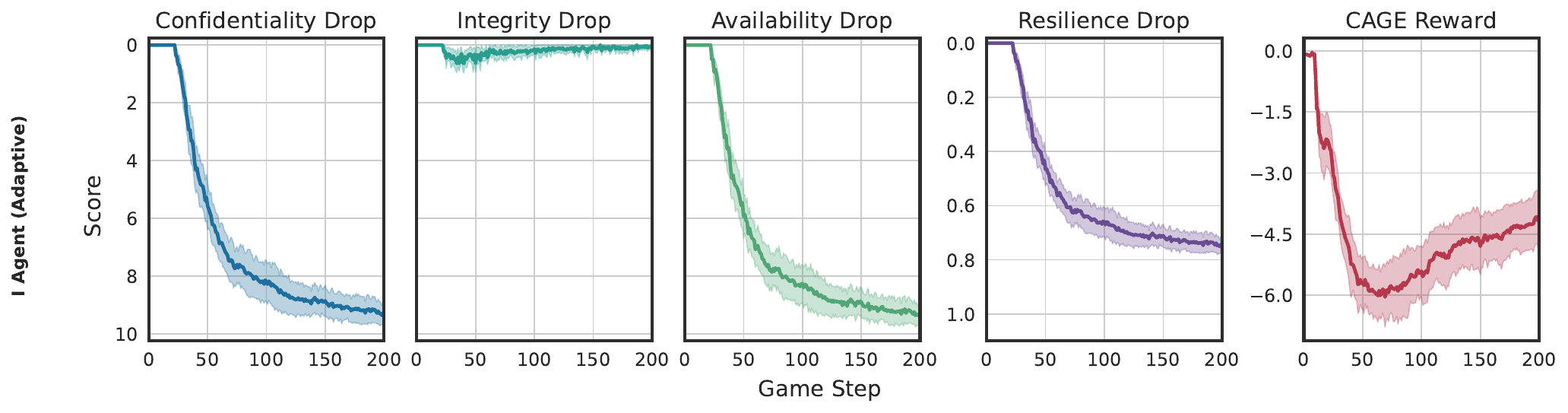}
    \caption{Comparing alignment approaches for the Integrity agent in the Base scenario (mean $\pm$ SEM, 100 episodes). Row 1: PPO baseline. Rows 2--3: test-time policy shaping (TT) with $\alpha=0.5$ and $\alpha=1.0$, respectively. Rows 4--5: training-time reward shaping (TR) with $\lambda=1.0$ and $\lambda=3.0$, respectively. Row 6: adaptive test-time policy shaping. Integrity shaping follows a similar pattern to Confidentiality: training-time $\lambda=3.0$ achieves the strongest suppression at Base scale, \rev{whereas test-time and adaptive policy shaping maintain more consistent alignment as network size increases}.}
    \label{fig:base_gameplay_example_I}
\end{figure*}

\begin{figure*}[h!]
    \centering
\includegraphics[trim={0 0.8cm 0 0},clip,width=0.75\textwidth]{figures/CAMLIS/appendix/base/Adaptive/200_ppo_agent.pdf}
\includegraphics[trim={0 0.8cm 0 0},clip,width=0.75\textwidth]{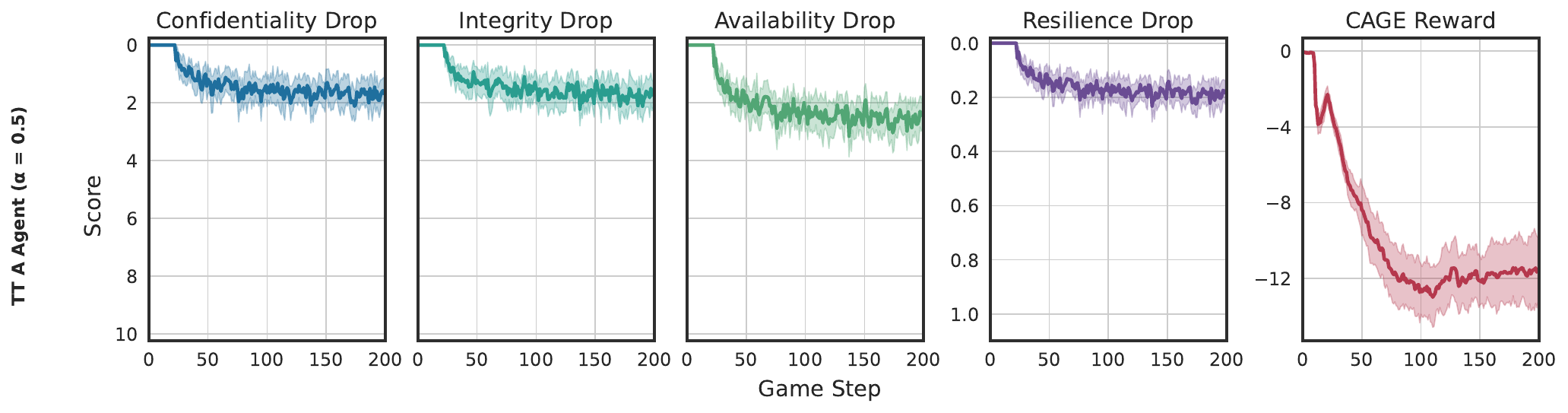}
\includegraphics[trim={0 0.8cm 0 0},clip,width=0.75\textwidth]{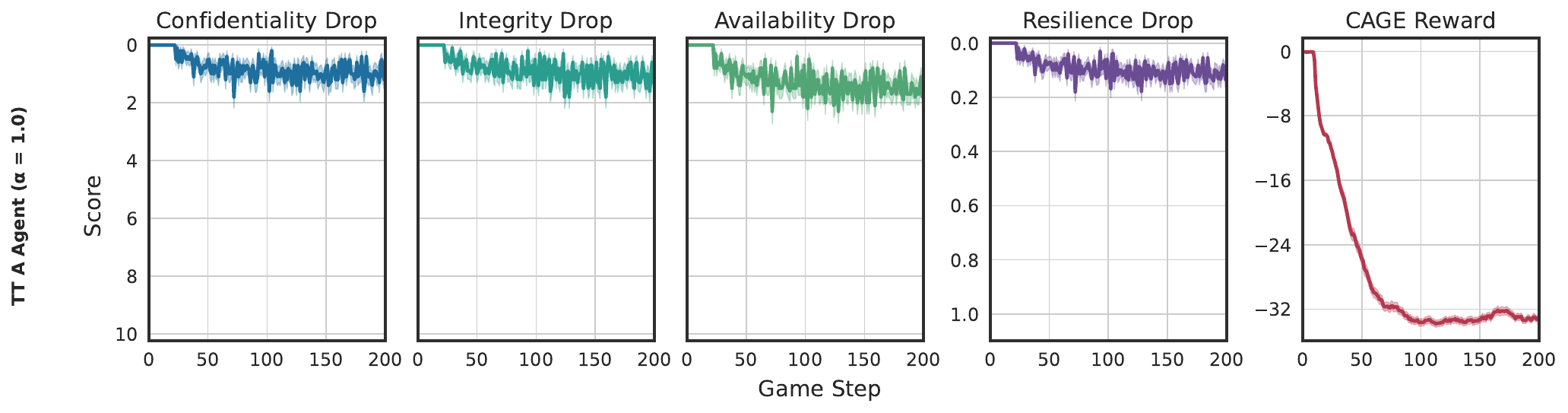}
\includegraphics[trim={0 0.8cm 0 0},clip,width=0.75\textwidth]{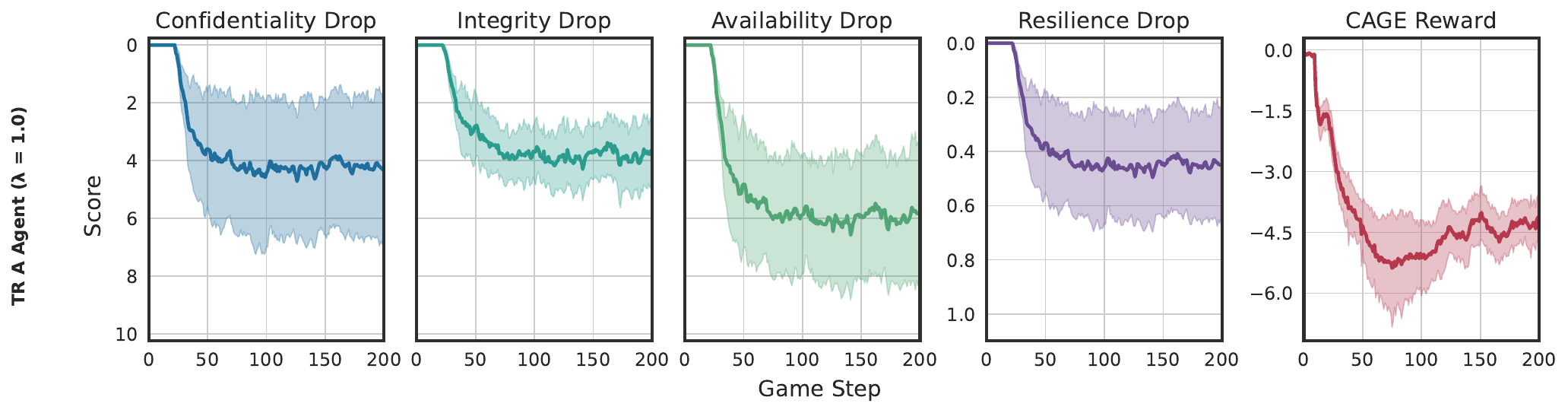}
\includegraphics[trim={0 0.8cm 0 0},clip,width=0.75\textwidth]{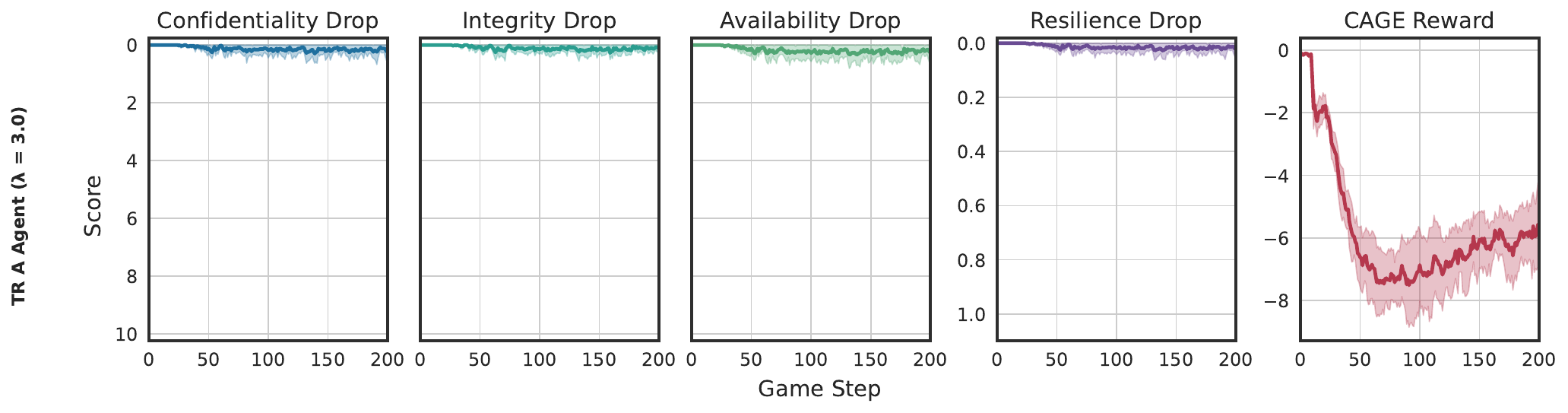}
\includegraphics[width=0.75\textwidth]{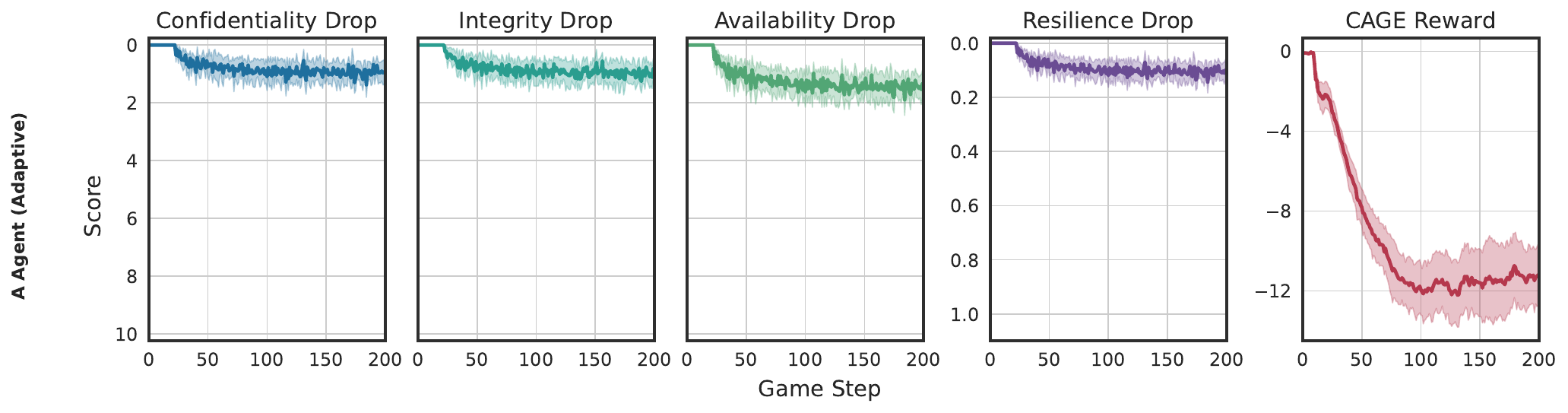}
    \caption{Comparing alignment approaches for the Availability agent in the Base scenario (mean $\pm$ SEM, 100 episodes). Row 1: PPO baseline. Rows 2--3: test-time policy shaping (TT) with $\alpha=0.5$ and $\alpha=1.0$, respectively. Rows 4--5: training-time reward shaping (TR) with $\lambda=1.0$ and $\lambda=3.0$, respectively. Row 6: adaptive test-time policy shaping. Availability shaping incurs a disproportionately higher reward cost compared to Confidentiality and Integrity, as all three CIA-critical servers contribute to Availability, making it inherently harder to protect without sacrificing task performance.}
    \label{fig:base_gameplay_example_A}
\end{figure*}

\begin{figure*}[t]
    \centering
        \includegraphics[width=0.8\textwidth]{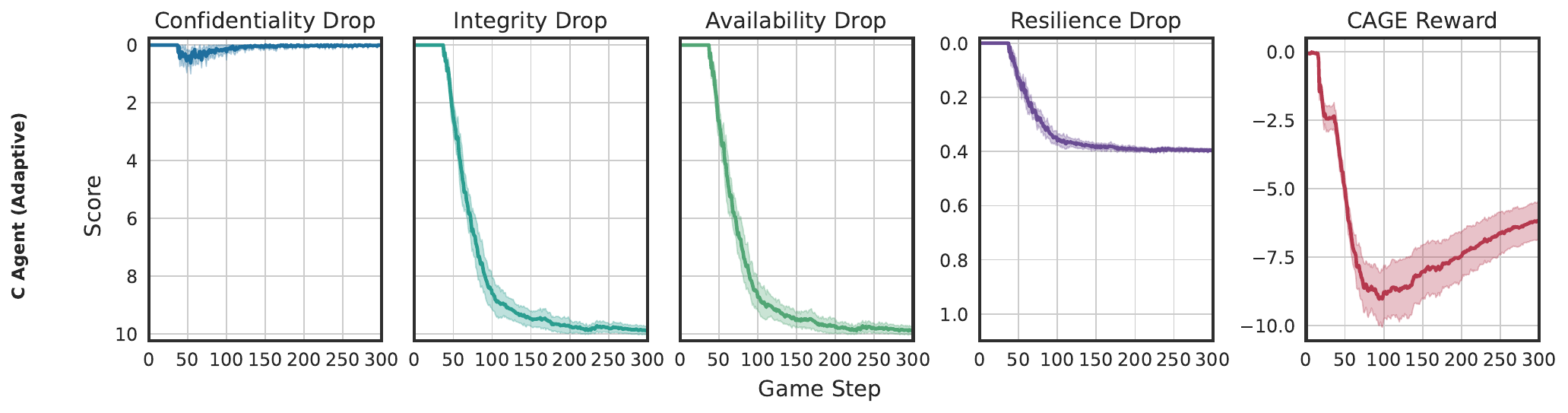}
        \includegraphics[width=0.8\textwidth]{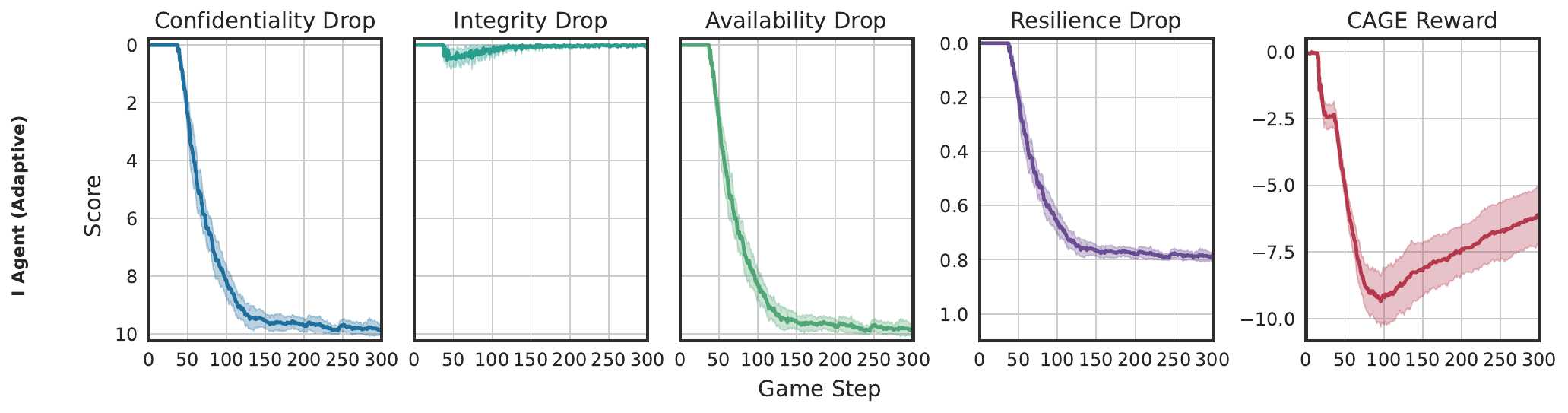}
        \includegraphics[width=0.8\textwidth]{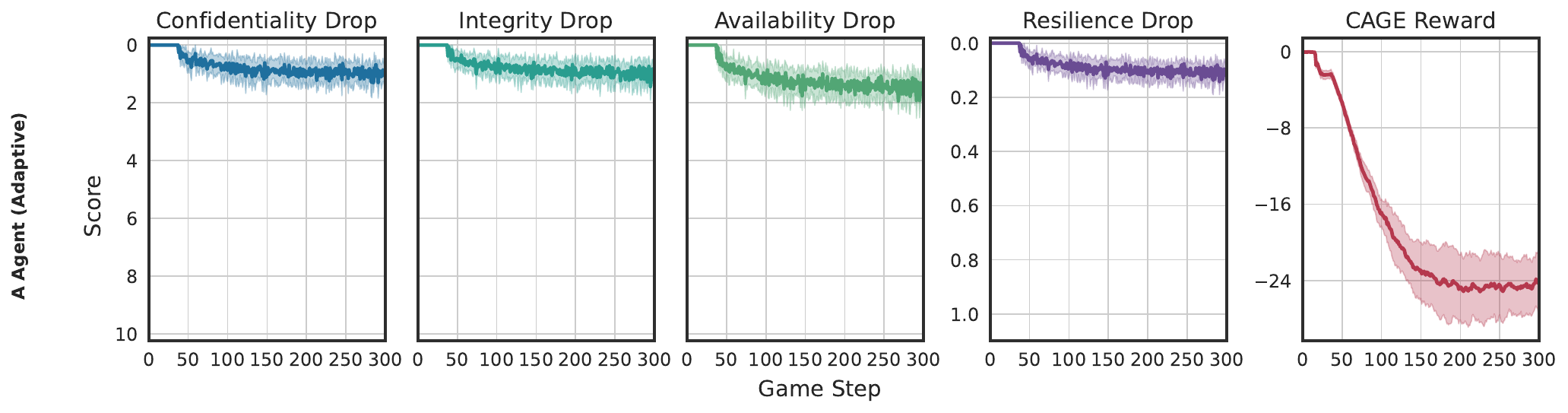}
    \caption{Adaptive policy shaping gameplay on the Hosts\_2 scenario (31 hosts, 300 steps).}
    \label{fig:gameplay_hosts2}
\end{figure*}

\begin{figure*}[h!]
    \centering

        \includegraphics[width=0.8\textwidth]{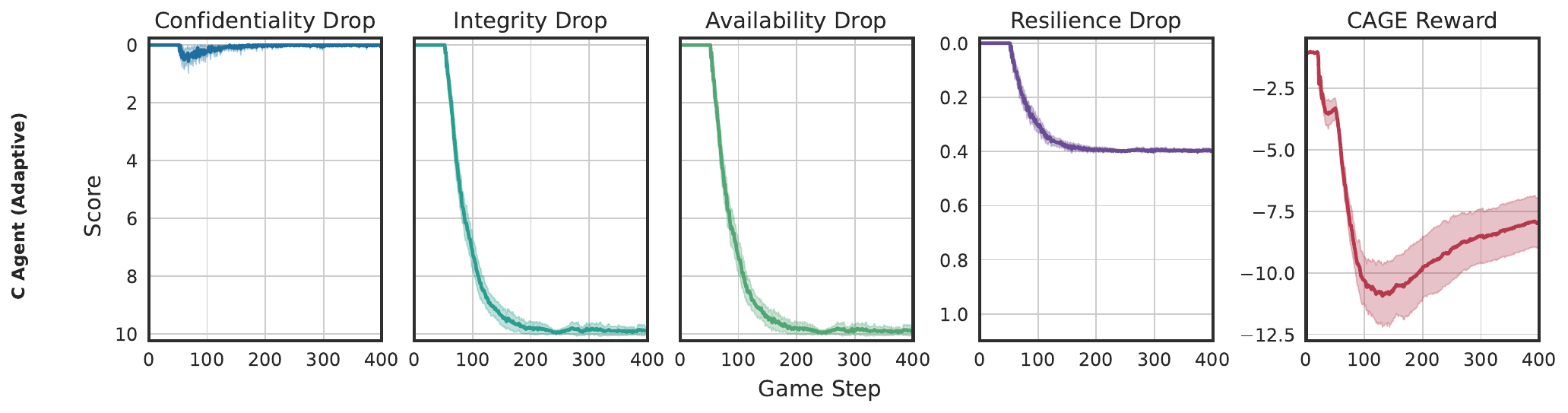}
        \includegraphics[width=0.8\textwidth]{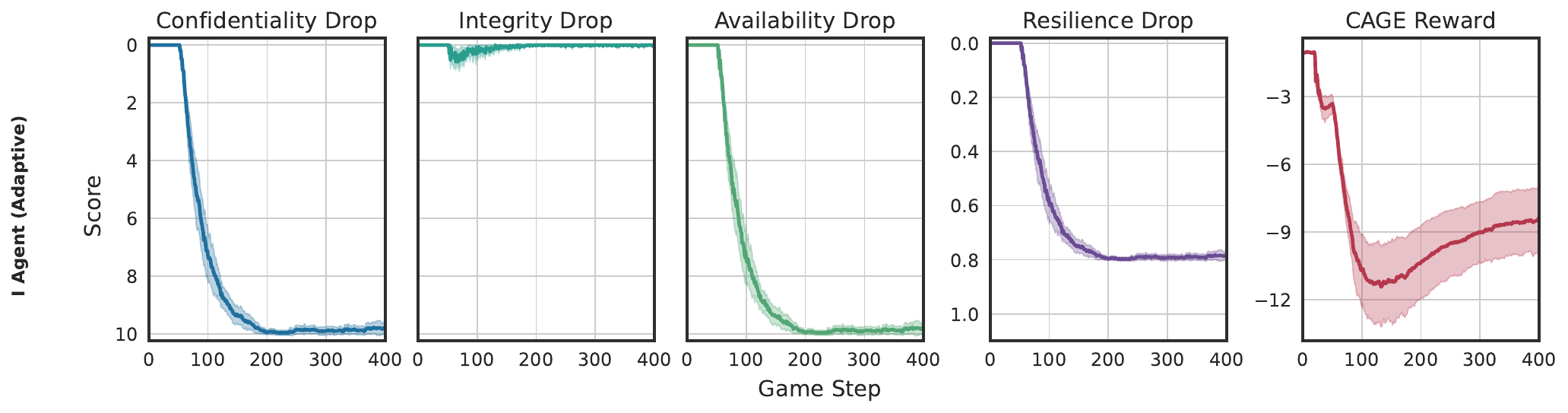}
        \includegraphics[width=0.8\textwidth]{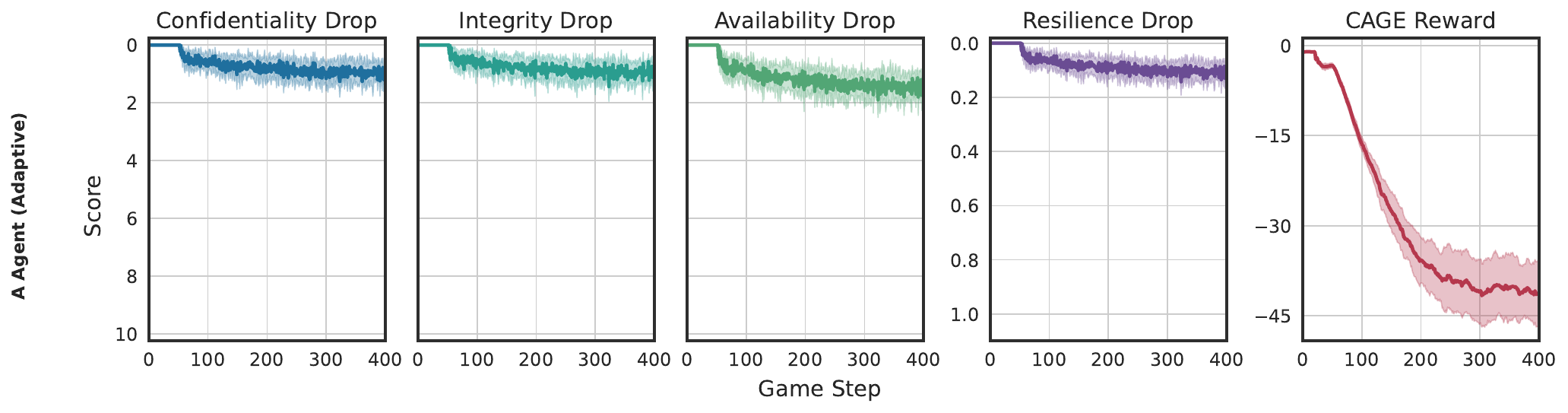}
            \caption{Adaptive policy shaping gameplay on the Hosts\_3 scenario (46 hosts, 400 steps).}
        \label{fig:gameplay_hosts3}
\end{figure*}

\begin{figure*}[h!]
    \centering

        \includegraphics[width=0.8\textwidth]{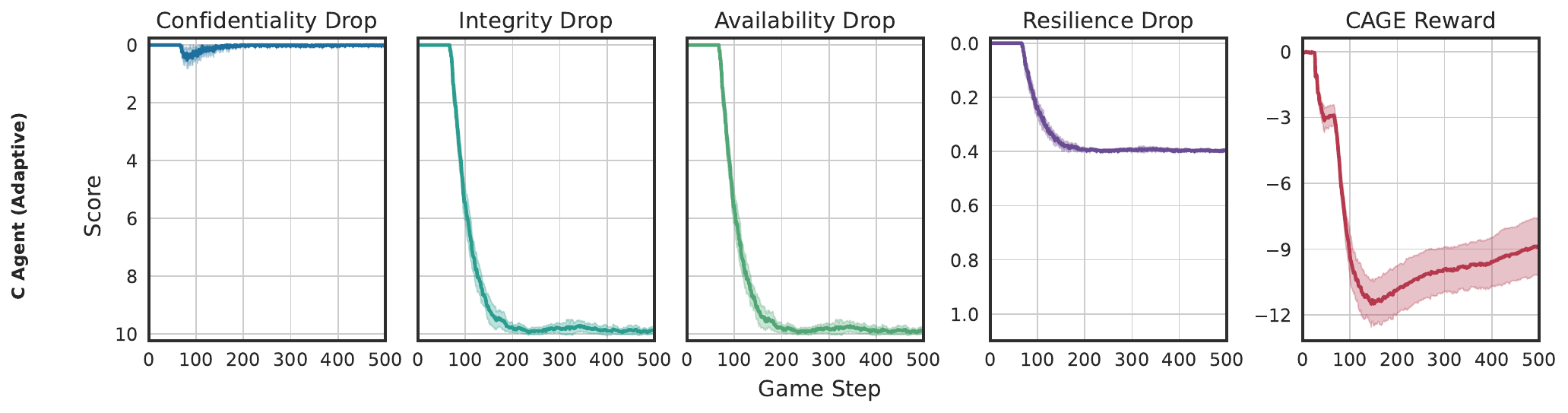}
        \includegraphics[width=0.8\textwidth]{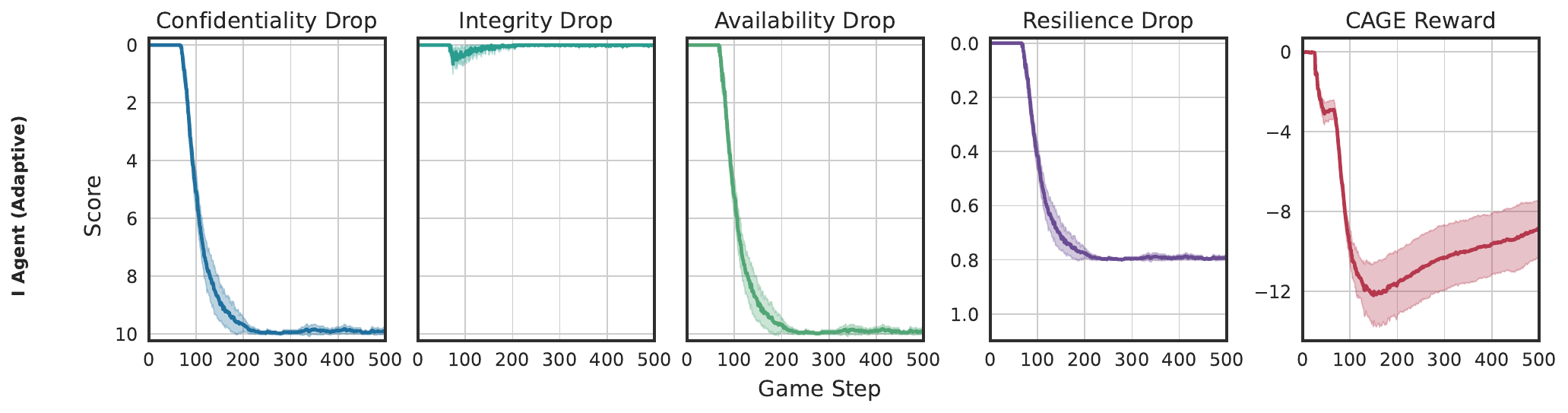}
        \includegraphics[width=0.8\textwidth]{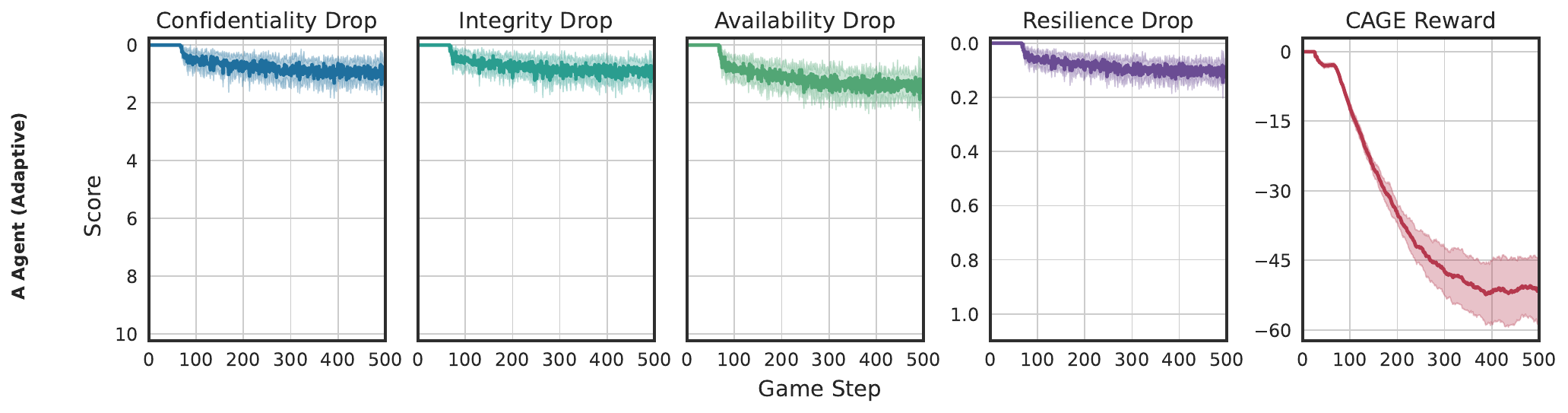}                    \caption{Adaptive policy shaping gameplay on the Hosts\_4 scenario (61 hosts, 500 steps).}
        \label{fig:gameplay_hosts4}
\vspace{-10pt}    
\end{figure*}

Figs.~\ref{fig:base_gameplay_example}--\ref{fig:base_gameplay_example_A} compare per-step confidentiality, integrity, and availability drop, resilience drop, and CAGE reward traces across PPO, test-time ($\alpha{=}0.5$ and $\alpha{=}1.0$), training-time ($\lambda{=}1.0$ and $\lambda{=}3.0$), and adaptive agents on the base scenario. Figs.~\ref{fig:gameplay_hosts2}, \ref{fig:gameplay_hosts3}, and \ref{fig:gameplay_hosts4} present per-step gameplay traces for the adaptive agent in \rev{the three larger network scenarios}, showing how CIA impact and resilience evolve over episode steps.



\end{document}

%% file: sections/abstract.tex
\begin{abstract}
Cybersecurity operators increasingly encounter sophisticated and fast-paced attacks, making autonomous cyber defense (ACD) a critical approach for supporting timely and effective decision-making. Reinforcement learning (RL)-based ACD agents have the ability to autonomously observe network states and take actions to defend and recover system assets, making them \rev{well suited for ACD}. However, these agents often face critical trade-offs, such as improving network security while inadvertently degrading service availability or violating other operational priorities such as confidentiality and integrity, raising the core challenge of aligning agent behavior with human-specified goals and intent. In this work, we study the problem of aligning RL-based ACD agents with human-specified operational objectives without requiring costly retraining. We formalize alignment for ACD as a policy shaping problem, where a pretrained RL decision-making policy can be adjusted at test time using predicted attributes over agent actions. We also propose an adaptive\del{ alignment} \rev{approach that dynamically controls the alignment--performance trade-off as operational priorities shift during an ongoing cyber attack}. We evaluate our method in a simulated cyber defense environment, CybORG, analyzing\del{ trade-offs between performance and alignment across multiple objectives}\rev{ performance across multiple alignment objectives} using the Confidentiality, Integrity, and Availability (CIA) information security triad. Our results show that adaptive test-time policy shaping can significantly improve alignment while preserving core cyber defense performance, offering a practical and scalable alternative to training-time alignment methods in RL. \rev{Our code is publicly available at \url{https://github.com/ITM-Kitware/alignment-cage-challenge-2}.}
\end{abstract}

%% file: sections/introduction.tex
\section{Introduction}
Modern-day cyber attacks have evolved in scale, sophistication, and impact, 
challenging traditional human-in-the-loop cybersecurity defenses and amplifying the risk of significant economic and societal harm \citep{lohn2023autonomous}. In response, a growing body of research has explored autonomous cyber defense (ACD), in which automated decision-making agents support the defense of large-scale networks against adversaries \citep{vyas2023automated}. ACD is typically formulated as a zero-sum adversarial game, in which a red agent models an attacker attempting to gain access to hosts and impact operational technologies, while a blue agent models a defender responsible for attack detection and recovery \citep{kiely2025cage}. This adversarial structure has made 
reinforcement learning (RL) approaches a natural fit for ACD, given the ability of agents to learn effective sequences of actions \proofread{with} limited prior information and their demonstrated success in other adversarial applications \citep{kiely2025cage,lohn2023autonomous}. Recent work has demonstrated the effectiveness of 
RL agents in simulated cyber environments \citep{kiely2025cage}, introduced benchmark cyber environments for agent development \citep{standen2021cyborg},
and explored \proofread{the} generalization of defensive strategies across network configurations and attackers \citep{wolk2022beyond}.

As cyber defense systems move beyond decision support to fully automated decision-making, the safety and downstream impact of ACD agents become critical concerns. Central to this challenge is \textit{alignment}: how an agent's actions correspond to human values, intentions, and priorities. Misalignment is likely when agents are trained to optimize a single reward or objective \citep{ji2023ai}. For example, a cyber defense agent may choose not to restore a compromised asset to avoid service downtime for end users, yet this decision can compromise confidentiality and increase the risk of data leakage. Similarly, aggressively isolating or rebuilding compromised hosts can limit further intrusion but disrupt availability for legitimate users. Such trade-offs are inherent to cyber defense and reflect broader challenges studied in other safety-critical domains, including clinical decision-making and medical triage \citep{hu-etal-2024-language}. However, alignment for ACD agents has received limited attention. Existing approaches typically optimize a single-objective environment reward and evaluate agents primarily on task performance, without explicitly incorporating human-defined operational \rev{objectives (e.g., confidentiality or integrity), while requiring retraining when objectives or network configurations change, thereby} limiting their adaptability in deployment.

\textbf{Our Contributions:} 
This paper addresses these challenges by formulating alignment of ACD agents as a test-time policy shaping problem, as illustrated in Fig. \ref{fig:overview}. The agent's \textit{policy} (i.e., its state-to-action decision rule represented as an action distribution) is reweighted at inference time and combined with an alignment-driven policy that promotes operator-specified objectives\del{ such as confidentiality, integrity, or availability}. By performing these adjustments at test time rather than during training, we avoid costly\del{ and time-consuming} retraining while enabling operational priorities to be dynamically adjusted as an incident unfolds. This approach also supports the incorporation of additional operational objectives and generalizes as network \rev{topologies} grow and attack strategies evolve. Overall, the contributions of our work are the following:

\begin{itemize}

\item \textbf{Adaptive test-time alignment of cyber defense agents:}
We formulate alignment as a test-time policy shaping framework over a fixed RL policy and investigate alignment to operational priorities of the Confidentiality, Integrity, and Availability (CIA) triad without costly retraining. This is in contrast to prior ACD work (e.g., \cite{kiely2025exploring,singhhierarchical}), which does not address agent alignment to human-specified objectives. We further address limitations of existing test-time policy shaping methods \citep{mujtaba2025aligning}, which rely on fixed strategies throughout gameplay (i.e., exclusively reward-driven or alignment-driven decisions), by introducing an adaptive alignment variant that dynamically switches between \proofread{the two} policies based on the observed environment state. This adaptive approach \del{more effectively balances the trade-off between reward and alignment when these objectives are in conflict}\rev{dynamically balances reward and alignment when these objectives conflict}.

\item \textbf{Alignment--performance trade-off analysis\del{ in cyber defense}:}
\rev{Although prior work on cyber resilience \citep{cadetquantitative} quantifies CIA-based operational objectives, it does not explicitly examine how these objectives trade off against agent reward. To address this gap,} we characterize the inherent trade-off between agent reward and alignment, showing that the cost structure is convex and scale-invariant and can vary with the red agent's attack strategy. Our empirical results \rev{suggest} that \del{an ACD agent with our adaptive approach offers a balanced and practical solution to capture this trade-off}\rev{adaptive policy shaping offers a practical approach to balancing these competing objectives}.

\item \textbf{Scalability and generalization to networks and attacks:}
To evaluate our approach, we train RL agents in the CybORG CAGE~2 environment \citep{kiely2023autonomous}, a popular testbed for developing and evaluating autonomous cyber defense agents \citep{wolk2022beyond}. This environment has been limited in scale compared to real-world networks, and prior work has not studied how alignment generalizes across network size or adversary strategy \citep{xie2025nash,kiely2025exploring}. Therefore, to investigate the generalization of our approach\del{ and the alignment--performance trade-off in cyber defense}, we extend this simulation environment to support configurable network topology generation and systematically vary network size, host importance weights, and red agent attack strategies in our evaluation. We demonstrate that test-time policy shaping generalizes across network sizes (16--61 hosts), attacker strategies, and alignment attributes, whereas training-time reward shaping collapses at larger network scales. This extended simulation environment can be found in our \rev{codebase \url{https://github.com/ITM-Kitware/alignment-cage-challenge-2}}.

\end{itemize}

%% file: sections/related_work.tex
\begin{figure}[t!]
\centering
\includegraphics[width=1\columnwidth]{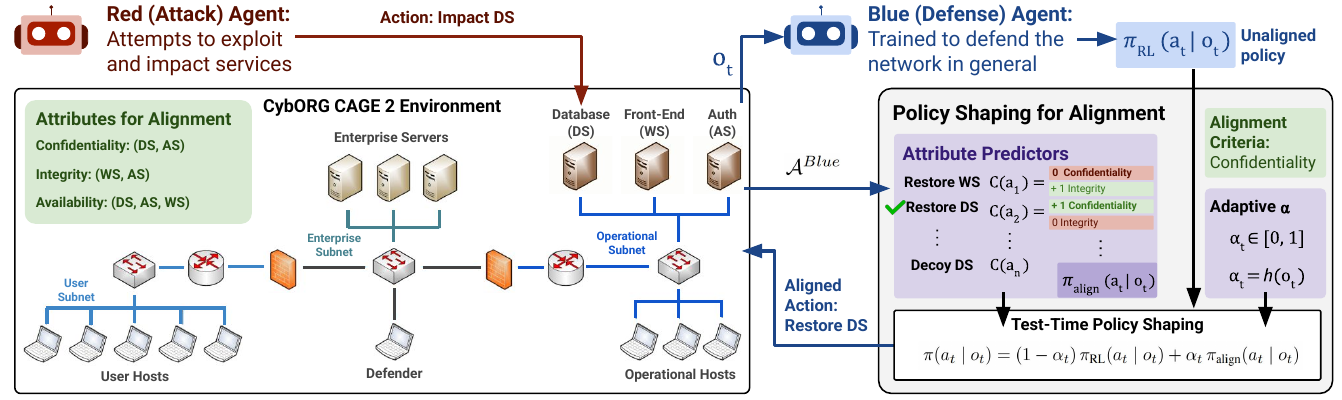}
\caption{In our modified CybORG CAGE 2 environment~\citep{kiely2023autonomous}, a red agent attacks network hosts while a blue agent defends. The blue agent's RL policy ($\pi_{\text{RL}}$) is trained to maximize 
reward but has no direct incentive to protect Confidentiality, Integrity, or Availability (CIA)-critical servers: Database (DS), Front-End (WS), and Auth (AS). 
To align an agent to these CIA operational priorities 
without retraining, we construct an alignment policy ($\pi_{\text{align}}$) using attribute predictors that score each action's CIA impact, and interpolate it with the RL policy at \proofread{test time} using policy shaping. 
}
\vspace{-10pt}
\label{fig:overview}
\end{figure}

\section{Related Work}

\textbf{RL for Autonomous Cyber Defense:} 
Early efforts to combat cyber attacks focused heavily on intrusion detection systems, which rely on techniques to identify malicious behavior by analyzing data such as application traces or network traffic flows to distinguish between normal and anomalous activity \citep{nguyen2021deep,buczak2015survey}. In contrast, ACD goes beyond detection by employing agents that can also take defensive actions and mitigate attacks in real time \citep{vyas2023automated}, enabling more effective responses to the growing volume and speed of modern cyber threats. Recently, ACD research has heavily relied on RL, particularly deep reinforcement learning (DRL), \proofread{in which} agents learn defensive or offensive strategies within adversarial simulation environments \citep{vyas2023automated,kiely2025cage,castro2025large,ridley2018machine}. RL is particularly well suited to cyber defense environments because it enables agents to learn sequential decision-making strategies under uncertainty, balancing long-term objectives while adapting to dynamic and adversarial network conditions. 

To support this line of research, several cyber defense environments have been developed, including Yawning Titan \citep{yawningtitan2022}, CyberBattleSim \citep{msft:cyberbattlesim}, and CybORG \citep{standen2021cyborg}\del{, which are designed for training and evaluating RL-based agents}. Prior work has explored a wide range of RL techniques in these environments, including deep Q-networks (DQN) \citep{mnih2015human}, proximal policy optimization (PPO) \citep{schulman2017proximal}, hierarchical RL \citep{singhhierarchical}, multi-agent RL \citep{kiely2025exploring}, and the co-training of attack and defense agents \citep{castro2025large}. \del{Additional studies have examined proactive and reactive defense strategies, as well as quantifying network resilience in ACD settings \mbox{\citep{cadetquantitative}}.} Despite this extensive body of work, little attention has been given to analyzing the behavior of ACD agents beyond task performance, particularly with respect to their alignment with human values and objectives.

\textbf{Human Value Alignment:} \textit{AI alignment} seeks to ensure that AI systems behave in accordance with human intentions, values, and goals \citep{ji2023ai}. Interest in alignment stems from documented cases of misaligned behavior, including violations of human values, manipulation, unintended resource acquisition, and power-seeking behavior \citep{ji2023ai,pan2023rewards}. Prior work has explored objectives such as robustness, interpretability, controllability, and adherence to human principles \citep{ji2023ai}. To date, much of this research has focused on large language models, using approaches including constitutional AI \citep{bai2022constitutional}, RL from human feedback \citep{bai2022training}, and RL from AI feedback \citep{rlaif2024}. Cyber defense introduces unique alignment challenges, as autonomous agents must balance security objectives against competing operational priorities such as service availability and mission continuity. \del{Alignment in ACD therefore extends beyond maximizing task performance and requires that defensive actions remain consistent with human-specified operational goals.}

\textbf{RL Alignment:}
RL agents optimize behavior through interaction with an environment and reward signals\del{, enabling effective long-term planning and adaptation based on real-time feedback} \citep{ates2020long}. However, prior work has shown that optimizing reward alone can lead RL agents to exhibit undesirable Machiavellian behaviors \citep{pan2023rewards,mujtaba2025aligning}. Approaches for aligning RL agents have therefore included reward modeling \citep{yu2025reward,zhang2025r1}, preference learning from human feedback \citep{hejna2024contrastive,poddar2024personalizing,hong2024adaptive}, and reward shaping or human-in-the-loop methods such as learning from demonstration \citep{chen2022human,wu2023toward,luo2025precise}. This differs from safe RL, which focuses on enforcing safety constraints and avoiding violations during optimization \citep{huang2024safety,huang2024human}, rather than learning or inferring the human preferences and values that define acceptable behavior. Value alignment presents a distinct challenge because human values are often heterogeneous, context-dependent, and difficult to specify compared to explicit safety constraints \citep{hu-etal-2024-language,mujtaba2025aligning}. While prior work has explored alignment for RL agents in a variety of domains, \rev{its application to adversarial and safety-critical settings such as ACD remains relatively underexplored.}

%% file: sections/methods.tex
\section{Methodology}\label{sec:methods}
\del{Autonomous cyber defense (ACD) }\rev{ACD} introduces trade-offs among multiple operational objectives whose relative priority depends on the attack scenario \citep{cadetquantitative}. To balance these trade-offs without costly retraining of defense agents, we propose an adaptive test-time policy shaping approach that dynamically switches the active decision-making policy of an ACD agent during incident response. We evaluate our approach in the CybORG CAGE 2 simulation environment, a widely used platform for ACD research \cite{kiely2023autonomous} (Sec. \ref{sec:problem_statement:simulation}), and study operational trade-offs through the \del{Confidentiality, Integrity, and Availability (CIA)}\rev{CIA triad} \del{attributes }(Sec. \ref{sec:problem_statement:alignment}). These attributes also serve as proxies for broader human-aligned objectives that may arise in cybersecurity, such as financial cost, mission impact, or service disruption, to which our approach can naturally extend. In the following sections, we also describe our baseline RL agents (Sec. \ref{sec:methods:agents}), present our test-time policy shaping approach for alignment (Sec. \ref{sec:methods:shaping}), present our adaptive policy shaping approach (Sec. \ref{sec:methods:adaptive}), compare our approach against training-time reward shaping baselines (Sec. \ref{sec:methods:training}), \del{describe how we will assess the generalizability of alignment under increasing network size and unseen adversaries}\rev{evaluate generalization across increasing network sizes and unseen adversaries} (Sec. \ref{sec:methods:generalization}), and \del{describe the metrics used for evaluation}\rev{define our evaluation metrics} (Sec. \ref{sec:methods:metrics}).

\subsection{Cyber Defense Simulation Environment}\label{sec:problem_statement:simulation}

We use the CybORG simulation environment \citep{standen2021cyborg}, focusing on the CAGE 2 scenario \citep{kiely2023autonomous}, which models a network of three subnets (Fig.~\ref{fig:overview}). We select CAGE 2 because it is a widely adopted benchmark for autonomous cyber defense and, through the modifications of \cite{cadetquantitative}, enables the study of resilience and CIA-driven operational trade-offs in a realistic attacker--defender setting. \del{We adopt the modified topology of \mbox{\cite{cadetquantitative}} for this environment, which introduces servers explicitly tied to the CIA alignment attributes.} \rev{We adopt this modified topology, which introduces servers tied to the CIA alignment attributes.} The red agent seeks to identify vulnerable hosts, escalate privileges, and execute \textit{impact} actions that cause service degradation. The blue agent defends through monitoring, removing malware, deploying decoys, and restoring compromised hosts \cite{kiely2023autonomous}.

We formalize the problem as a partially observable Markov decision process (POMDP) \citep{sutton1998introduction}, \proofread{treating the attacker as part of the environment dynamics. The POMDP is defined by} $\langle \mathcal{S}, \mathcal{A}, \mathcal{T}, \mathcal{O}, r^{\text{CAGE}}, b_0 \rangle$ \citep{kiely2023autonomous,le2025learning}. Here, $\mathcal{S}$ is the state space representing the network configuration, $\mathcal{A}$ the set of defensive actions (determined by the number of hosts and per-host actions), $\mathcal{T}(s' \mid s, a)$ the transition function that implicitly captures agent interactions, and $\mathcal{O}(o \mid s, a)$ the observation function providing the blue agent with partial information about host status and detected red agent sessions. The initial belief $b_0$ encodes uncertainty over the starting configuration. The base blue agent reward in CAGE, $r^{\text{CAGE}}(s, a) \in \mathbb{R}$, reflects penalties incurred from red agent compromise and impact actions, as well as blue agent restore actions (i.e., restoring a machine to its original state, which causes one \proofread{time step} of service disruption):
\begin{equation}
\label{cage-reward}
r^{\text{CAGE}}_t = -0.1(\text{host}_{\text{R}} - 1) - \text{server}_{\text{R}} - \text{B}_{\text{Restore}} - 10(\text{R}_{\text{Impact}}).
\end{equation}
\proofread{Here,} $\text{host}_{\text{R}}$ and $\text{server}_{\text{R}}$ denote hosts and servers compromised by the red agent $R$, respectively, and $\text{B}_{\text{Restore}}$ is an indicator equal to 1 if \proofread{the blue agent} $B$ has run the restore action on any host or server, and 0 otherwise, and $\text{R}_{\text{Impact}}$ is an indicator equal to 1 if \proofread{the red agent} has run the impact action on any host or server \citep{kiely2023autonomous}. The maximum possible reward is 0. Importantly, this reward function excludes hosts and servers associated with the CIA attributes, resulting in an environment that enables the study of the trade-off between operational alignment objectives and the original performance-driven reward.

\input{tables/scenarios}

\subsection{Confidentiality, Integrity, and Availability (CIA) Attribute Scores}\label{sec:problem_statement:alignment}
 Following \cite{cadetquantitative}, \rev{we focus on the CIA attributes as domain-specific alignment objectives, which can also be used to collectively characterize a system's \textit{cyber resilience}---its ability to resist, absorb, and recover from compromise. We use preservation of these attributes as a domain-specific proxy for broader alignment objectives, grounding our formulation in prior cyber-resilience research.} Confidentiality (C) captures unauthorized access to sensitive data, Integrity (I) reflects unauthorized modification of data, and Availability (A) denotes disruption of services for legitimate users. These attributes are mapped to specific network assets: $\{\text{AS}, \text{DS}\}$ for Confidentiality, $\{\text{AS}, \text{WS}\}$ for Integrity, and $\{\text{AS}, \text{DS}, \text{WS}\}$ for Availability. To quantify each attribute's score, we adapt the network resilience formulation of \cite{cadetquantitative}, which defines an indicator $N_j(t) = 1$ if an adversarial impact on critical asset $j$ is successful at time $t$, and $N_j(t) = 0$ otherwise. Each asset $j$ then carries a cost $\text{cost}(i, j)$ representing the penalty to attribute $i \in \{\text{C}, \text{I}, \text{A}\}$. We then compute an \textit{attribute drop score} over a time window $\Delta t$ as the \rev{cumulative} impact on assets associated with each attribute; for example, the confidentiality drop is defined as
\begin{equation}
\label{attribute-score}
C(B, R, \Delta t) = \sum_{j \in \{\text{AS}, \text{DS}\}}
\left( \sum_{t \in \Delta t} N_j(t) \right) \cdot \text{cost}(\text{C}, j),
\end{equation}
where $B$ and $R$ denote the blue and red agent actions, respectively. The \textit{resilience drop} is then defined as the weighted sum of the $\{\text{C}, \text{I}, \text{A}\}$ drop scores. Formally, this is defined as the following:
\begin{equation}
\label{resilience-drop}
\Delta Resilience Drop(B, R, \Delta t) = \sum_{i \in \{\text{C}, \text{I}, \text{A}\}} w_i \sum_{j \in \{\text{AS}, \text{DS}, \text{WS}\}}
\left( \sum_{t \in \Delta t} N_j(t) \right) \cdot \text{cost}(i, j).
\end{equation}

Whereas \proofread{\citet{cadetquantitative}} use an indicator of a successful adversarial impact to measure resilience loss, we invert this indicator to measure the ``damage prevented'' as a proxy for alignment with the CIA attributes. Specifically, we set $N_j(t) = 1$ if an adversarial impact on critical asset $j$ is prevented or resolved at time $t$ (e.g., the blue agent restores a compromised host), and $N_j(t) = 0$ otherwise (e.g., a host was not impacted by the red agent, or the red agent compromises the host and the blue agent fails to restore it). Under this formulation, actions that prevent or remediate red impacts to an asset, such as restoring a compromised host, contribute positively to the corresponding CIA attribute. Similarly, an agent may remove an active malicious process to keep a host online rather than fully restoring it from a clean image, thereby preserving availability. However, this strategy requires monitoring the host to identify the malicious process before it can be removed. 

We adopt a multi-objective formulation treating alignment to these attributes and the CAGE reward as distinct objectives. \proofread{To this end}, since defending CIA-associated assets may divert attention from other parts of the network, we construct a modified CAGE scenario that permits red impacts on all hosts and removes CIA-critical hosts from the blue agent reward calculation entirely. This decoupling prevents unintended correlation between alignment and reward, more closely reflecting real-world settings where cyber defenders must balance multiple competing priorities.

\subsection{Autonomous Cyber Defense RL Agents}\label{sec:methods:agents}
Following prior work~\citep{kiely2023autonomous}, we train blue agents in the \rev{CybORG} CAGE 2 environment using the Proximal Policy Optimization (PPO) algorithm \citep{schulman2017proximal}. The actor and critic are represented by two feedforward neural networks, \proofread{each with two hidden layers of 256 units}. Each blue agent is trained against a red agent in the base scenario described in Sec. \ref{sec:problem_statement:alignment}, \proofread{using 200-step games in which the blue and red agents each take an action at every step. We train each agent for} 500k \proofread{steps with a} learning rate of 0.003, \proofread{a} discount factor of 0.99, \proofread{and a} mini-batch size of 1024 for the SGD optimizer. Agent code, including seeds to reproduce results, \rev{is} provided in our codebase. 

CybORG is packaged with several different heuristic red and blue agents, which we rely upon for training and evaluation. The base Red Meander agent has no knowledge of the network topology and randomly navigates the network, looks for vulnerabilities, and attempts to exploit hosts on the \proofread{network until} it is able to impact a host with the operational technology (OT) service present. For our purpose of training blue agents, we create a new Red Combination agent which switches randomly with probability $p=0.5$ between a policy to impact CAGE-specific hosts (i.e., hosts not tied to CIA attributes) and \proofread{a policy to impact} alignment-specific CIA hosts (i.e., the database, front-end, and authentication servers). Other agents available for testing are described in Sec. \ref{sec:methods:generalization}.

\subsection{Alignment with Test-Time Policy Shaping}
\label{sec:methods:shaping}

Unlike prior work that measures agent success primarily through environment reward or resilience metrics \citep{kiely2023autonomous, cadetquantitative}, we explicitly model operational objectives as alignment signals that may conflict with reward maximization. Our approach builds on the \textit{test-time policy shaping} framework of \citet{mujtaba2025aligning}, \proofread{in which} the resulting agent policy $\pi$ is a convex combination of two or more action distributions. Such ensemble-based methods have been shown to improve robustness in decision-making settings \citep{liublending,czarnecki2020real,mujtaba2025aligning}. 

Given a \proofread{pretrained} RL policy $\pi_{RL}$ and an alignment-driven policy $\pi_{align}$, the shaped policy is defined as:
\begin{equation}
 {\pi(a_t \mid o_t)} = (1 - \alpha_t)\, \mathbf{\pi}_{\text{RL}}(a_t \mid o_t) + \alpha_t\, \mathbf{\pi}_{\text{align}}(a_t \mid o_t)
\end{equation}
\proofread{where,} at time $t$, $a_t \in \mathcal{A}^{Blue}$ is a blue agent action and $o_t \in \mathcal{O}$ is the blue agent's partial observation of the network. The interpolation parameter $\alpha_t \in [0, 1]$ controls the degree of attribute-based policy shaping: $\alpha_t = 1$ yields full control by the alignment policy, while $\alpha_t = 0$ recovers the base RL agent. Varying $\alpha_t$ traces the Pareto front between reward maximization and alignment adherence.

To construct the alignment policy $\pi_{align}$, we compute per-action CIA attribute scores over the blue agent action space $\mathcal{A}^{\text{Blue}}$. Let $\{k_i\}_{i=1}^{M}$ denote the alignment-relevant attributes, with $k_i  \in \{\text{C}, \text{I}, \text{A}\}$. For an action $a_t \in \mathcal{A}^{\text{Blue}}$ under observation $o_t$, we form an attribute-score vector $\mathbf{z}(a_t \mid o_t) \in \mathbb{R}^{M}$ whose $i$-th component measures the one-step, cost-weighted contribution of action $a_t$ toward attribute $k_i$, computed only over the assets associated with that attribute (e.g., $\{\mathrm{AS}, \mathrm{DS}\}$ for confidentiality).

We define the target (ideal) vector $\boldsymbol{\tau} \in \mathbb{R}^{M}$ \rev{as} the maximum attainable score for attribute $k_i$ (i.e., by performing actions that maximize that particular attribute).
Each action is scored by its Euclidean distance to the target:
$d(a_t \mid o_t) = \sqrt{\sum_{i=1}^{M} \left(z_i(a_t \mid o_t) - \tau_i\right)^2}.$
The alignment policy is then defined as a softmax over negative distances:
\begin{equation}
\pi_{\text{align}}(a_t \mid o_t) =
\frac{\exp(-d(a_t \mid o_t))}{\sum_{a' \in \mathcal{A}^{\text{Blue}}} \exp(-d(a' \mid o_t))}
\end{equation}
This formulation assigns higher probability to actions closer to the desired attribute target while maintaining exploration among similarly aligned actions. For example, under a confidentiality objective, actions that \proofread{restore the authentication and database servers or remove malicious processes from them} may receive higher probability, whereas under an availability objective, actions that maintain critical hosts may be preferred. Although we demonstrate this approach with an on-policy RL agent, the interpolation framework applies equally to off-policy settings \citep{mujtaba2025aligning}.

\subsection{Alignment with Adaptive Policy Shaping}\label{sec:methods:adaptive}
During deployment, the \proofread{relative priorities of} reward maximization and alignment adherence may change as the state of the network evolves. Rather than relying on a fixed interpolation parameter $\alpha$, we propose an \textit{adaptive test-time policy shaping} mechanism that sets $\alpha_t$ as a deterministic, rule-based function of the current observation: $\alpha_t = h(o_t)$. Given the blue agent's partial observation,
$h(o_t)$ flags the state as unsafe if any alignment-relevant host is currently impacted; if so, we set $\alpha_t = 1$ to prioritize the alignment policy, and otherwise set $\alpha_t = 0$ to revert to the reward-maximizing policy. \rev{We use this binary rule as an interpretable, training-free mechanism that isolates the benefit of state-dependent policy shaping without introducing an additional learned controller or associated training assumptions.} We \rev{also} note that, while this effectively amounts to \textit{policy switching}, our approach can generalize to other $h(o_t)$ functions that directly predict context-dependent $\alpha_t$ values from observations. The resulting action distribution is then computed via the same convex combination of the reward- and alignment-driven action probabilities, with $\alpha_t$ determined by $h(o_t)$.

\subsection{Alignment with Training-Time Reward Shaping}
\label{sec:methods:training}
As a baseline comparison to our test-time policy shaping approach, we investigate alignment using \textit{training-time reward shaping}. This approach modifies the reward function to include an alignment-dependent bonus term, weighted by a coefficient $\lambda$ relative to the original CAGE reward, which encourages actions consistent with a specified alignment objective during training. We adopt a bonus-based formulation rather than a weighted sum of independent reward components, since alignment-relevant events (i.e., undesirable impacts to CIA-critical hosts) occur infrequently during training, making an additive bonus more effective. The shaped reward function then becomes:
\begin{equation}
r_t = \tilde{r}_t^{\text{CAGE}} + \lambda \, r_t^{\text{align}}(k),
\qquad
k \in \{C, I, A\},
\end{equation}
where $\tilde{r}_t^{\text{CAGE}}$ is the normalized CAGE environment reward for the blue agent (Eq.~\ref{cage-reward}), and $r_t^{\text{align}}(k)$ is an alignment bonus derived from the attribute score (Eq.~\ref{attribute-score}), which is 1 if the selected action improves the attribute $k$ and 0 otherwise. For a given target attribute $k$, the alignment term is computed from the observed red and blue agent actions and
observations at time $t$ and scaled by the maximum attainable attribute score.
To improve training stability and convergence, we normalize the CAGE reward signal using a warm-up period of ten episodes to estimate the running mean and standard deviation, after which rewards are normalized and clipped before being combined with the alignment bonus. PPO agents trained with reward shaping use the same hyperparameters as the baseline.

\subsection{Generalization and Scalability}\label{sec:methods:generalization}
A central challenge in autonomous cyber defense is generalization from simulated environments to real-world environments, which often involve larger network topologies, more diverse adversary behaviors, and shifting defensive priorities. To study this in a more controlled setting, we develop a network topology generator for CybORG that enables systematic variation in subnet size and reward weighting. We construct four scenarios -- \textbf{base} (the original CAGE~2 topology with 16 hosts), \textbf{hosts-2} (31 hosts), \textbf{hosts-3} (46 hosts), and \textbf{hosts-4} (61 hosts) -- each scaling the number of hosts per subnet while proportionally increasing the training budget and the number of steps per episode (see Tab.~\ref{tbl:network_demographics}). As network size grows, the number of CIA-critical hosts increases, making sustained adversarial impact more likely and cyber defense more challenging.

We evaluate the robustness of blue agents against several different
red agents: a \textbf{Meander CIA Agent (M-CIA)} that randomly navigates the network while targeting CIA-critical hosts; a \textbf{Meander CAGE Agent (M-CAGE)} that targets hosts associated with the CAGE reward but not CIA attributes; and \textbf{B-line Agents (Bline-C/I/A)} that possess prior knowledge of the network topology and can consistently target assets associated with a specific attribute, resulting in more persistent and challenging attack patterns. 

Together, these scenarios and red agent types allow us to assess whether alignment-driven defenses generalize across network scales and remain robust under diverse adversarial conditions. Our new CAGE 2 environment is also available for use in our code repository.

\subsection{Evaluation Metrics}\label{sec:methods:metrics}
For all experiments, we evaluate both agent performance and operational alignment. The CAGE environment \textbf{reward} (Eq.~\ref{cage-reward}) measures overall defensive performance, where the maximum achievable reward is 0. We also define an attribute-specific \textbf{alignment score}, which measures the fraction of decision steps in which the blue agent selects an action aligned with a target operational attribute $k_i \in \{\mathrm{C}, \mathrm{I}, \mathrm{A}\}$. Formally, let $a_t$ denote the action selected by the blue agent at time step $t$, and let $\mathcal{A}_{k_i}(o_t) \subseteq \mathcal{A}^{\mathrm{Blue}}$ denote the set of actions that are aligned with attribute $k_i$ under observation $o_t$. The alignment score for attribute $k_i$ is then defined as
$
\frac{1}{T}
\sum_{t=1}^{T}
\mathbf{1}
\left[
a_t \in \mathcal{A}_{k_i}(o_t)
\right],
$ where $T$ is the episode length and $\mathbf{1}[\cdot]$ is the indicator function \rev{of a selected aligned action}.

We \del{additionally}\rev{also} measure\del{the} overall network \textbf{resilience drop} (Sec.~\ref{sec:problem_statement:alignment}) using \del{a weighting scheme from}\rev{the weighting scheme of} \rev{\citet{cadetquantitative}}\del{, with $w_C = 0.6$ and $w_I = w_A = 0.2$, and assume equal costs for all assets}\rev{ ($w_C = 0.6$, $w_I = w_A = 0.2$), and assume equal asset costs}. \del{While these parameters are application-dependent}\rev{Although these parameters} are application-dependent, our focus is on evaluating alignment methods rather than \del{selecting}\rev{tuning} application-specific weights. Agents that successfully prioritize assets associated with the target operational objective \del{produce}\rev{achieve} lower \del{impact and }resilience-drop scores.

%% file: tables/scenarios.tex
\begin{table}[t!]
    \centering
    \resizebox{0.5\textwidth}{!}{%
    \begin{tabular}{l|cccc}
        \hline
        & \textbf{Base} & \textbf{Hosts\_2} & \textbf{Hosts\_3} & \textbf{Hosts\_4} \\
        \hline
        Total Hosts          & 16  & 31  & 46  & 61  \\
        \quad User Subnet    & 5   & 10  & 15  & 20  \\
        \quad Enterprise Subnet & 4 & 7   & 10  & 13  \\
        \quad Operational Subnet & 7 & 14 & 21  & 28  \\
        \hline
        CIA-Critical Hosts   & 3   & 6   & 9   & 12  \\
        \quad Auth Servers    & 1   & 2   & 3   & 4   \\
        \quad Database Servers & 1  & 2   & 3   & 4   \\
        \quad Front-End Servers   & 1   & 2   & 3   & 4   \\
        \hline
        Steps / Episode      & 200 & 300 & 400 & 500 \\
        Action Space $|\mathcal{A}|$ & 178 & 343 & 508 & 673 \\
        \hline
    \end{tabular}}
    \vspace{-5pt}
    \caption{Network topologies across scenarios of increasing scale. CIA-critical hosts are scored for Confidentiality, Integrity, and Availability impact. The enterprise subnet includes the defender host.}
    \vspace{-10pt}
    \label{tbl:network_demographics}
\end{table}

%% file: sections/results.tex
\section{Results and Discussion}
We \del{present results of our alignment approach on ACD agents across four experiments}\rev{evaluate our alignment approaches for ACD agents across four sets of experiments}. First, \del{in Sec. \ref{sec:results:baseline}, we evaluate baseline blue agent performance against a range of red agent adversaries. We compare the blue monitor (\textbf{BM}) and \textbf{PPO} blue agents against red agents targeting CIA operational priorities (\textbf{M-CIA}), targeted adversaries focused on confidentiality, integrity, or availability (\textbf{Bline-C}, \textbf{Bline-I}, and \textbf{Bline-A}), and adversaries that randomly target CAGE hosts without considering operational priorities (\textbf{M-CAGE}). This experiment evaluates the ability of blue agents to defend the network under adversaries with differing operational objectives.}\rev{we establish baseline performance of blue monitor (\textbf{BM}) and \textbf{PPO} agents against adversaries with differing operational objectives (Sec. \ref{sec:results:baseline}).} Second, \del{in Sec. \ref{sec:results:tr_alignment}, we align agents at training-time (\textbf{RL-$\lambda$}), with varying values of $\lambda$ reward bonus, and in Sec. \ref{sec:results:alignment}, evaluate our test-time policy shaping approach (\textbf{RL-$\alpha$}) under varying values of $\alpha$ and compare it to the standard training-time alignment approach, assessing effectiveness in adapting to attacks with differing operational priorities without costly retraining.}\rev{we compare training-time reward shaping (\textbf{RL-$\lambda$}) and test-time policy shaping (\textbf{RL-$\alpha$}) across varying $\lambda$ and $\alpha$ values, assessing their effectiveness in balancing alignment and reward (Secs. \ref{sec:results:tr_alignment} and \ref{sec:results:alignment}).} Third, \del{in Sec. \ref{sec:results:adaptive} we study adaptive selection, which addresses a limitation of test-time policy shaping, where improved alignment may come at the cost of reduced competence or reward.}\rev{we evaluate adaptive policy shaping, which dynamically adjusts alignment priorities to mitigate the trade-off between alignment and reward (Sec. \ref{sec:results:adaptive}).} Finally, \del{in Sec. \ref{sec:results:generalization} we perform a comprehensive analysis of all three alignment approaches, evaluating their ability to maintain competence and alignment under held-out adversaries (e.g., novel attack strategies not seen during training) and as network size increases. This is important in real-world cyber defense settings, where agents must generalize to unseen network configurations and evolving adversarial strategies due to continuously changing infrastructure and threat landscapes.}\rev{we assess the generalization and scalability of all alignment approaches under held-out adversaries and increasing network sizes (Sec. \ref{sec:results:generalization}).}

All experiments \del{were run}\rev{ran} on a single NVIDIA RTX A6000 GPU. For each experiment, five blue agents were trained using different seeds and evaluated on held-out evaluation seeds. The results presented are aggregated across these five trained agents. \rev{Seeds and parameters are available in our code repository, with RL training curves provided in the Appendix.}

\subsection{Baseline Blue Agent Performance}\label{sec:results:baseline}
\begin{figure}[t!]
\centering
\includegraphics[width=\columnwidth]{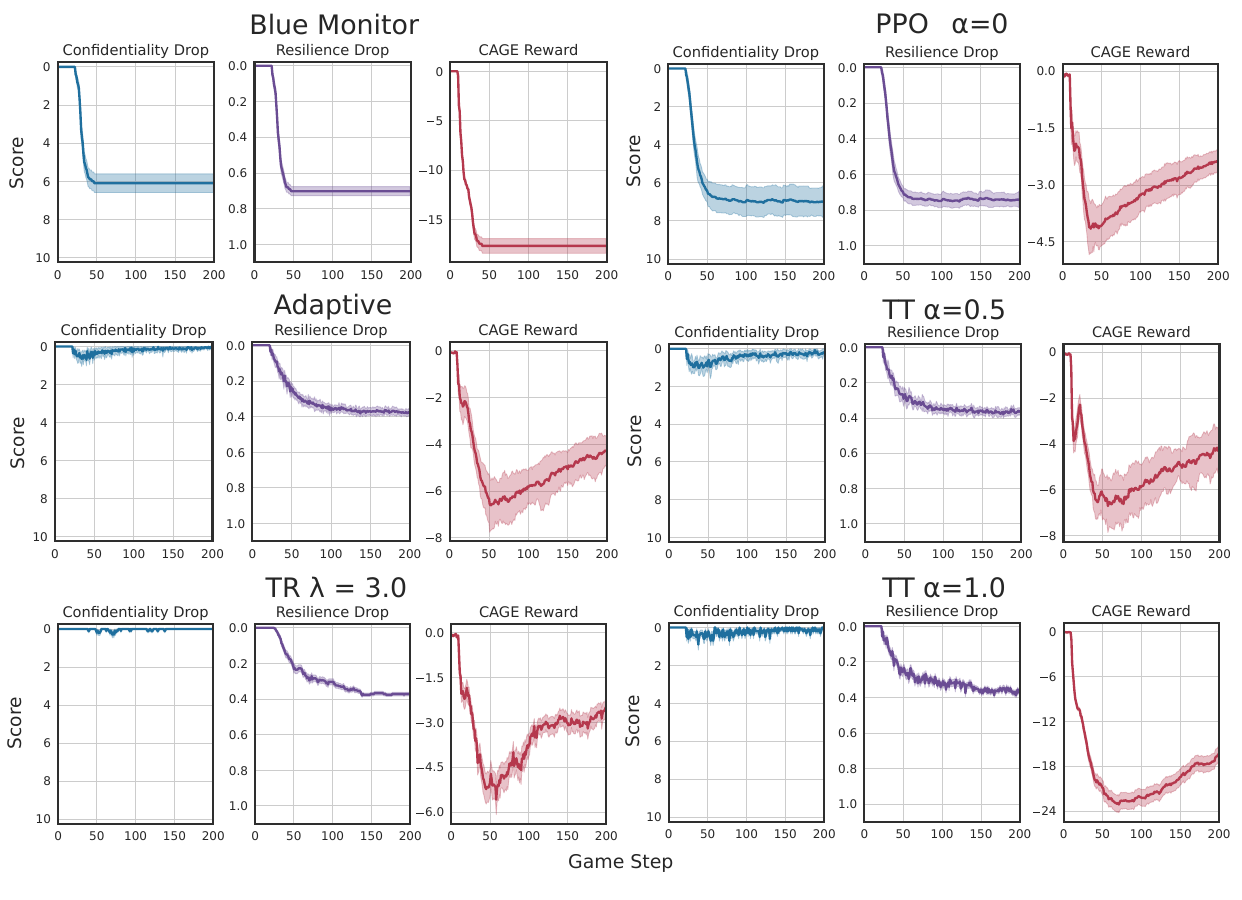}
\vspace{-10pt}
\caption{Per-step comparison of blue agent performance against the red M-CIA agent on the base scenario\del{(16 hosts)}. Each panel shows the Confidentiality drop, Resilience drop, and CAGE reward (mean $\pm$ SEM across 100 episodes, each 200 steps); \rev{values closer to zero indicate better performance for all metrics}. \textbf{Blue monitor} \rev{provides a lower bound on performance across metrics; in this baseline, the agent only monitors the network and takes no defensive actions}. \textbf{PPO} is the unaligned baseline agent. \textbf{TT ($\alpha$=0.5)} and \textbf{TT ($\alpha$=1.0)} \protect\rev{apply} test-time policy shaping to interpolate between the Confidentiality-aligned policy and the fixed PPO policy. \textbf{TR ($\lambda$=3.0)} is trained with a Confidentiality alignment bonus in the reward. \textbf{Adaptive} dynamically switches between the fixed PPO and Confidentiality-aligned policies based on agent observations. All three alignment approaches substantially reduce Confidentiality drop relative to the unaligned PPO baseline. 
}
\label{fig:all_approaches}
\end{figure}


The results across different blue agents and red adversaries on the base scenario (see Tab.~\ref{tbl:network_demographics}) are presented in Fig. \ref{fig:all_approaches} for the M-CIA agent that targets at random a CIA asset. The Blue Monitor\del{ (\textbf{BM})} agent provides a lower bound on the reward, as it only monitors the network and takes no defensive actions. The PPO agent is trained to maximize the CAGE reward and achieves the highest reward by protecting the operational server, but fails to defend CIA-critical hosts. This results in a significant resilience drop\del{and impacts to Confidentiality, Integrity, and Availability assets}, comparable to the BM agent. This gap between ACD agent performance and CIA impact motivates the use of the three alignment approaches\del{ evaluated here}: test-time policy shaping (Sec.~\ref{sec:methods:shaping}), training-time reward shaping (Sec.~\ref{sec:methods:training}), and adaptive policy shaping (Sec.~\ref{sec:methods:adaptive}). These approaches extend the baseline PPO policy to reduce CIA impact while preserving as much reward advantage as possible.

\subsection{Training-Time Reward Shaping}\label{sec:results:tr_alignment}

A standard approach for steering agents toward different objectives is to shape the reward signal during training. In alignment settings, this can consist of providing a ``bonus'' for taking actions that align with specific attributes or human-specified objectives \citep{hendrycks2021would}. Therefore, we first study the application of alignment at training time to provide a comparison with our test-time approaches. The primary disadvantage of training-time alignment is that, for every change in operational priorities or human values, the agent must be retrained, which can be both time-consuming and computationally expensive.

Results for training-time alignment\del{ of blue agents} are illustrated in Fig. \ref{fig:comparison_lambda}, with additional results and training curves provided in the Appendix. We compare varying values of $\lambda$\del{, which controls the alignment bonus applied during training,} across different operational priorities\del{. We}\rev{, and} observe that as $\lambda$ increases, the agent learns to better protect CIA-aligned assets while maintaining relatively consistent reward performance.\del{ Fig. \ref{fig:comparison_lambda} shows the effect of varying $\lambda$ for the blue agent against the M-CIA red agent.} The primary challenge with this approach is that a new agent must be retrained whenever operational priorities change, which can become increasingly time-consuming as network size grows (as illustrated later in Sec. \ref{sec:results:generalization}). Furthermore, training on a specific adversary \rev{(e.g., M-CIA)} can lead to poorer performance against unseen adversaries \rev{(e.g., M-CAGE, which targets CAGE-only assets)}; this is illustrated in \rev{Appendix Tabs. S1--S4}, where in the case of attribute \rev{``A''} (i.e., all CIA assets are prioritized) and $\lambda=3.0$, the training-time agent's reward is worse than the baseline PPO agent across all scenarios. This reflects the trade-off associated with prioritizing the defense of one set of operational objectives over another.

\begin{figure}[t!]
\centering
\includegraphics[width=\columnwidth]{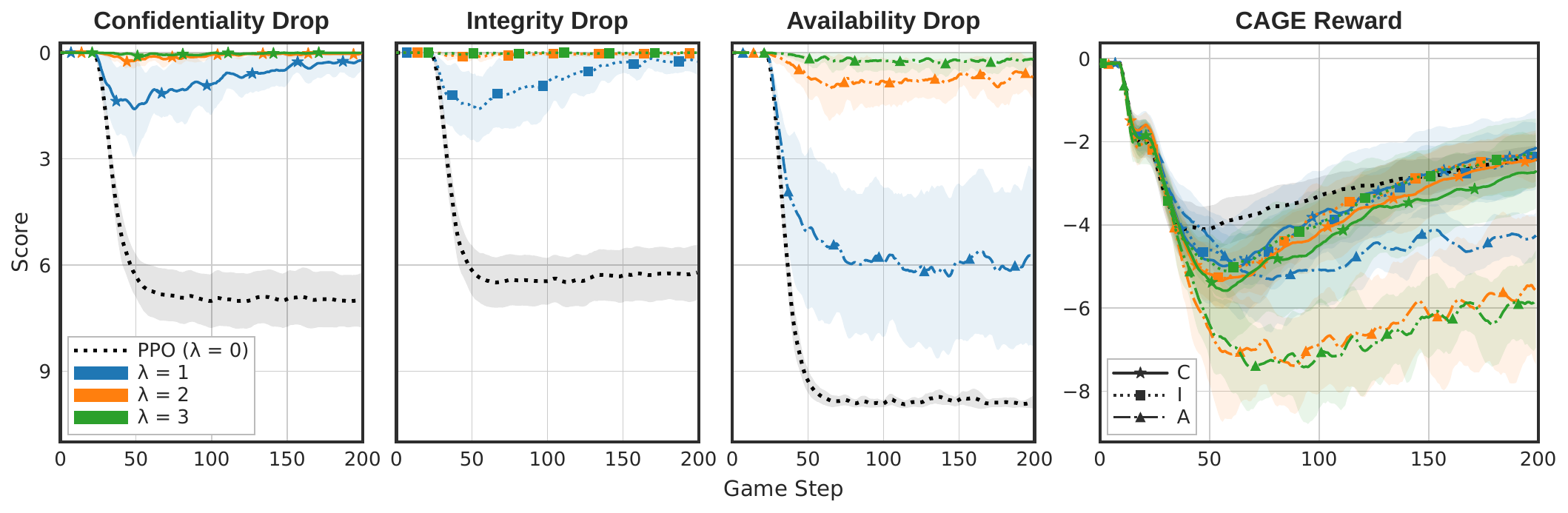}
\vspace{-10pt}
\caption{Comparison of scores (mean $\pm$ SEM across 100 episodes, \rev{against M-CIA red}) as \rev{the alignment bonus $\lambda$} is increased in training-time reward shaping for each CIA attribute. Across all attributes, increasing $\lambda$ \rev{improves CIA drop scores and reward}.}
\label{fig:comparison_lambda}
\end{figure}

\subsection{Test-Time Policy Shaping to Operational Priorities}\label{sec:results:alignment}

\begin{figure}[t!]
\centering
\includegraphics[width=\columnwidth]{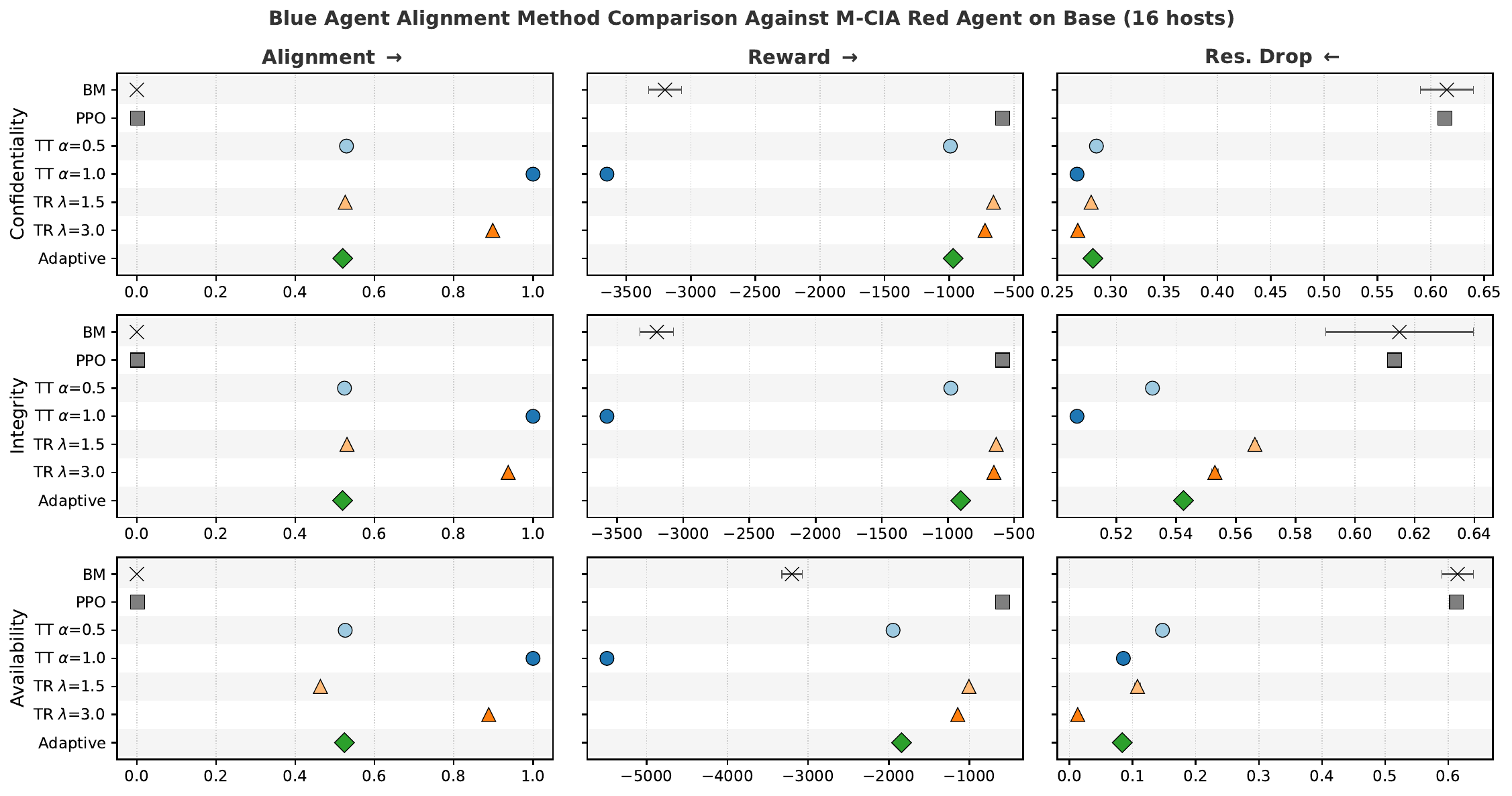}
\vspace{-10pt}
\caption{\del{Overall}Alignment score, reward, and resilience drop\del{,} of alignment approaches against the M-CIA red agent\del{(that randomly targets any one CIA asset)}. Each point shows \rev{the mean performance of the method listed on the left, averaged across five seeds with 100 episodes of 200 steps each}. Horizontal bars represent \proofread{the} standard deviation across seeds; \rev{this variation is very small for all approaches except BM}\del{and is therefore not visible in the plot}.}
\label{fig:overall}
\end{figure}

\begin{figure}[t!]
\centering
\includegraphics[width=\columnwidth]{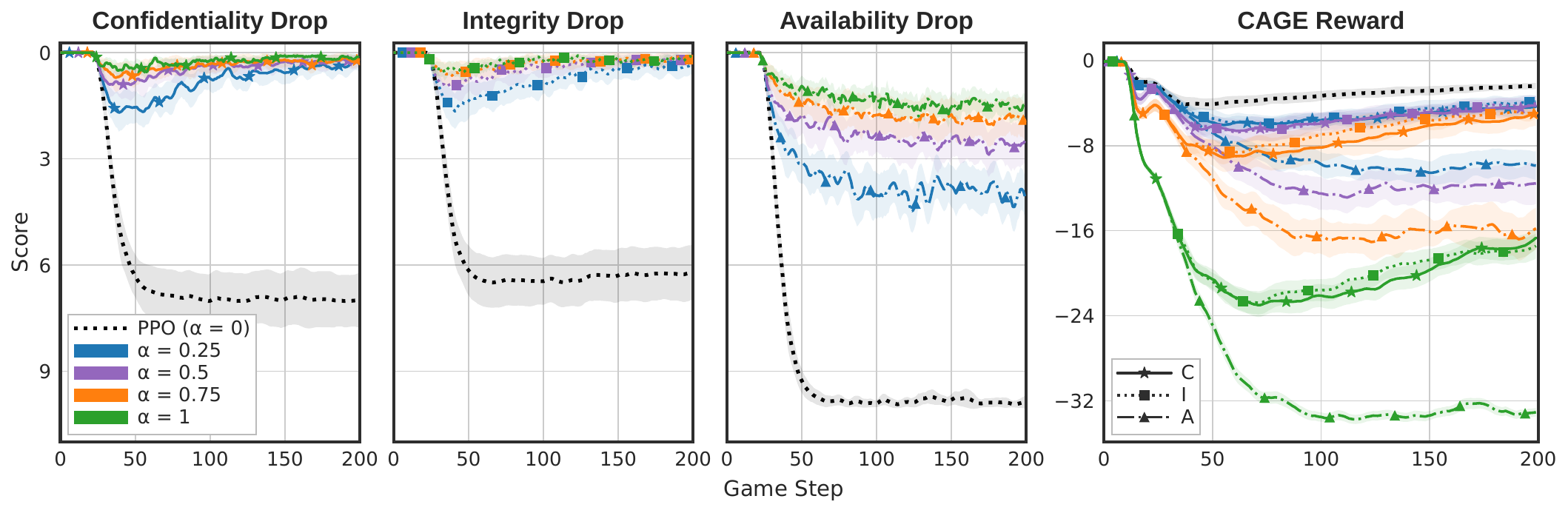}
\vspace{-15pt}
\caption{Comparison of scores (mean $\pm$ SEM across 100 episodes, \rev{against M-CIA red}) as $\alpha$ \rev{varies} in test-time policy shaping \rev{for each CIA attribute}. As $\alpha$ increases, \rev{alignment improves substantially, but reward decreases}.}
\label{fig:200_alpha_comparison}
\end{figure}

To \del{provide an adaptable and efficient approach for steering the operational priorities of}\rev{efficiently and flexibly steer the operational priorities of} ACD agents, we propose test-time policy shaping and \del{adaptive test-time policy shaping}\rev{its adaptive variant} as alternatives to \del{the standard training-time approach}\rev{training-time alignment}. \del{Results for test-time policy shaping across red agents are illustrated in Fig. \ref{fig:overall}, with additional results provided in the Appendix, alongside training-time alignment and baseline agents.}\rev{Fig.~\ref{fig:overall} compares test-time policy shaping with training-time alignment and baseline agents against the M-CIA red agent, with additional results in the Appendix.} We \del{compare}\rev{consider} two settings: $\alpha=0.5$, which \del{weighs the original PPO reward-driven policy and alignment-driven policy equally}\rev{assigns equal weight to the original PPO reward-driven and alignment-driven policies}, and $\alpha=1.0$, which \del{relies entirely on}\rev{uses only} the alignment-driven policy. Across all red agent types, \del{we find that} test-time policy shaping ($\alpha=1.0$) \del{produces the most aligned agents while also reducing}\rev{achieves the strongest alignment while reducing} attribute-associated impacts and resilience drops, \del{but at the cost of reward}\rev{albeit at the cost of lower reward}. \del{This is illustrated in Fig. \ref{fig:200_alpha_comparison}, where increasing values of $\alpha$ place greater weight on CIA-aligned actions, resulting in a corresponding decrease in agent reward.}\rev{Fig.~\ref{fig:200_alpha_comparison} illustrates this trade-off: increasing $\alpha$ places greater weight on CIA-aligned actions while decreasing agent reward.} \del{In contrast}\rev{By contrast}, training-time reward shaping ($\lambda=3.0$) \del{yields agents that often achieve}\rev{often achieves} the best impact scores, with smaller \del{decreases}\rev{reductions} in reward and alignment scores. Intermediate test-time ($\alpha=0.5$) and training-time ($\lambda=1.5$) settings \del{provide a smooth interpolation}\rev{interpolate} between \del{the }baseline PPO and fully attribute-aligned policies. \del{This trade-off motivates an adaptive approach that can alternate between these conflicting objectives.}

\begin{figure}[t!]
\centering
\includegraphics[width=.48\columnwidth]{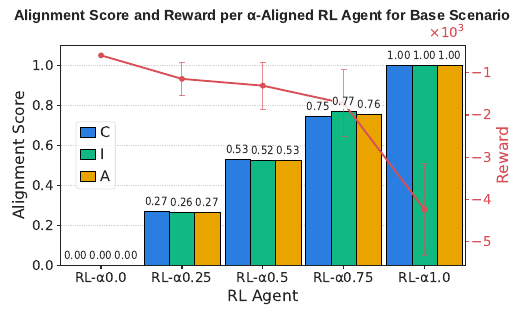}
\includegraphics[width=.48\columnwidth]{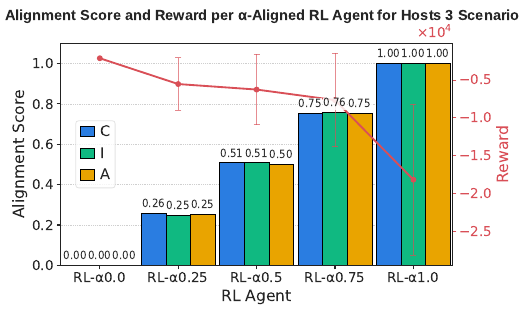}
\includegraphics[width=.48\columnwidth]{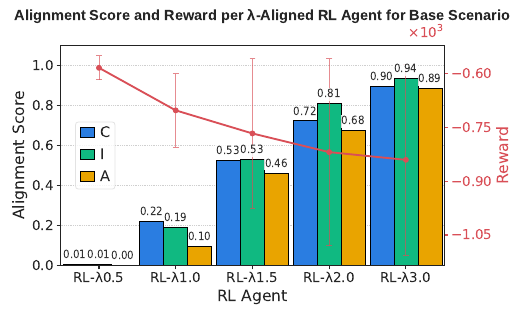}
\includegraphics[width=.48\columnwidth]{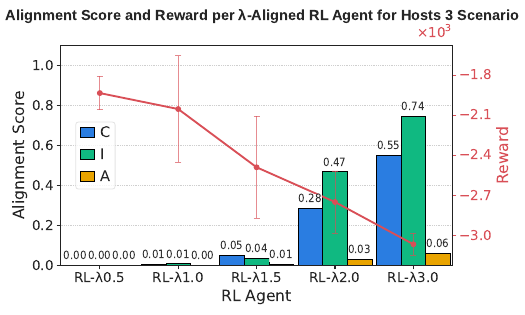}
\vspace{-10pt}
\caption{Test-time policy shaping \rev{(top row)} and training-time reward shaping \rev{(bottom row)} results for a blue agent \del{playing against}\rev{facing} the red M-CIA agent \del{on}\proofread{in} the base \rev{(left column)} \del{scenario and $Hosts_3$ scenario}\rev{and $Hosts\_3$ \rev{(right column)} scenarios}. \del{There is a fundamental trade-off}\rev{The results show a trade-off} between alignment scores and CAGE reward \del{as a function of $\alpha$ and $\lambda$}\rev{as $\alpha$ and $\lambda$ vary}. Colored bars show alignment scores (left axis) for the Confidentiality (C), Integrity (I), and Availability (A) attributes; the red line shows the average CAGE reward (right axis). \rev{For test-time policy shaping, alignment scores consistently improve as network size and $\alpha$ increase. In contrast, training-time reward shaping maintains reward alongside alignment as $\lambda$ increases, but its alignment scores collapse as network size grows.}}
\label{fig:base_CIA_alignment}
\end{figure}

Fig.~\mbox{\ref{fig:base_CIA_alignment}} (top row) \rev{further} quantifies the test-time alignment-reward trade-off \rev{as $\alpha$ varies}. Alignment scores for all three CIA attributes scale nearly linearly with $\alpha$ and are balanced across CIA at every network scale tested. While the magnitude of the reward changes with scale (reflecting the increased difficulty in defending larger networks), the reward cost remains convex across scales: moderate $\alpha$ values offer favorable cost-benefit ratios, while $\alpha = 1.0$ guarantees full compliance at a steeper penalty, allowing human operators to tune $\alpha$ at deployment time without retraining. \rev{These trade-offs motivate an adaptive approach that can alternate between the competing objectives.}

\subsection{Adaptive Alignment under Performance Trade-offs}
\label{sec:results:adaptive}

\begin{figure}[t!]
\centering
\includegraphics[width=.9\columnwidth]{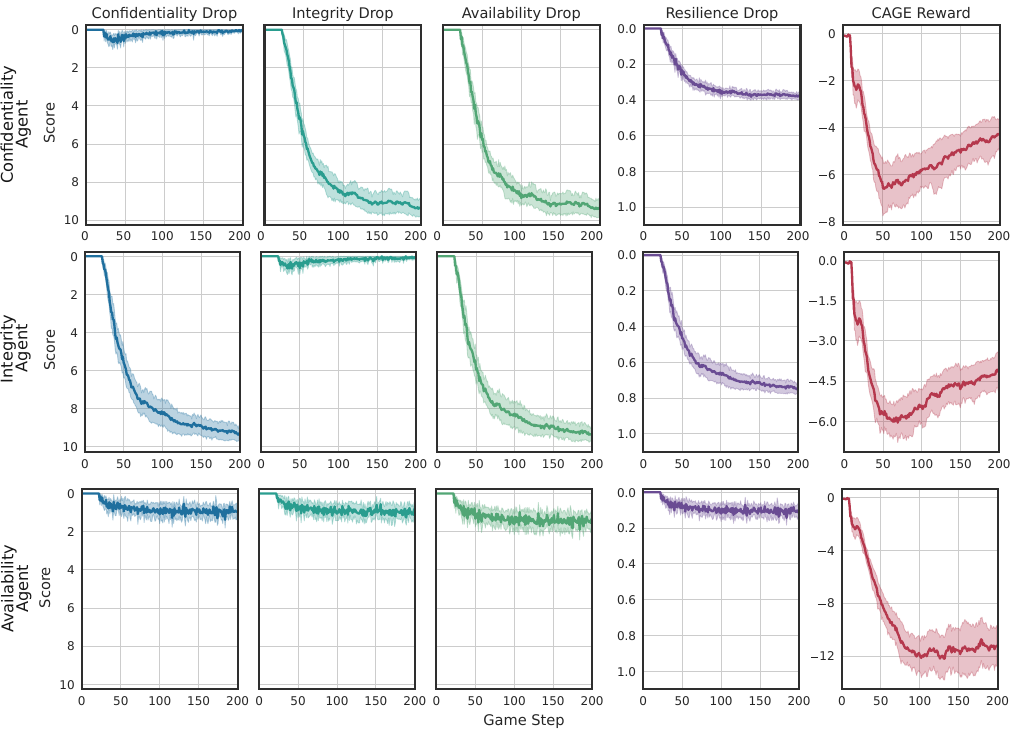}
\vspace{-10pt}
\caption{Adaptive blue agent performance against the M-CIA red agent in the base scenario (mean $\pm$ SEM across 100 episodes). Rows show Confidentiality-, Integrity-, and Availability-aligned agents using adaptive $\alpha$. C- and I-agents reduce their target CIA drops to near zero within 20 steps; suppression by the A-agent is more \rev{variable and less complete}.}
\label{fig:adaptive_cia}
\end{figure}

\rev{Our adaptive policy shaping approach is demonstrated in Fig. \ref{fig:adaptive_cia} for all CIA attributes.} This agent uses current observations to infer the red M-CIA agent's attack strategy and \rev{switches between defending} the operational server (preserving CAGE reward) \rev{and protecting} specific CIA-critical hosts (reducing CIA resilience drops). \rev{Across these objectives, the} \proofread{drops} in Confidentiality and Integrity \proofread{are} \rev{mitigated}, whereas the Availability drop is more difficult for the agent to counter, likely because Availability is linked to all three CIA-critical hosts\del{ (DS, WS, and AS)}, thereby expanding the scope of cyber defense required.

\rev{To place these results in context, Fig.~\ref{fig:overall} compares the adaptive approach with the other alignment approaches across CIA objectives and reward against the M-CIA red agent, while Fig.~\ref{fig:all_approaches} illustrates the alignment--reward trade-off for Confidentiality in the base scenario.} \del{While the baseline PPO agent}\rev{The unaligned PPO baseline} achieves the highest reward\del{, it exhibits}\rev{ but has} substantial drops in both Confidentiality and overall alignment \del{because}\rev{as} the red agent \del{successfully }compromises CIA-critical hosts. Test-time policy shaping, in contrast, progressively reduces Confidentiality drop as $\alpha$ increases, \del{although this improvement comes at the cost of reduced reward. Training-time reward shaping produces comparable suppression of Confidentiality impact while maintaining higher reward performance, having internalized both operational objectives during training.}\rev{at the cost of reward, while training-time reward shaping achieves comparable suppression with higher reward.} \rev{The adaptive agent achieves suppression of Confidentiality loss similar to test-time policy shaping with $\alpha=1.0$ by activating alignment only after detecting compromise of Confidentiality-associated assets.} \del{The adaptive agent achieves suppression similar to test-time policy shaping with $\alpha=1.0$ by activating alignment only after compromise of Confidentiality-associated assets is detected.} \del{Analysis of the adaptive agent's behavior}\rev{Behavioral analysis} shows that \del{it}\rev{the adaptive agent} disrupts the red agent's attack chain by prioritizing \del{the defense of }CIA-critical hosts, preventing the lateral movement \del{required}\rev{needed} to reach downstream operational servers. Although all three alignment approaches substantially reduce Confidentiality impact relative to the PPO baseline, resilience is not fully recovered because the red agent can still compromise Integrity- and Availability-associated assets that are not explicitly defended under a Confidentiality objective. Additional alignment results are in the Appendix.

\begin{figure}[t]
\centering
\includegraphics[width=.4\columnwidth]{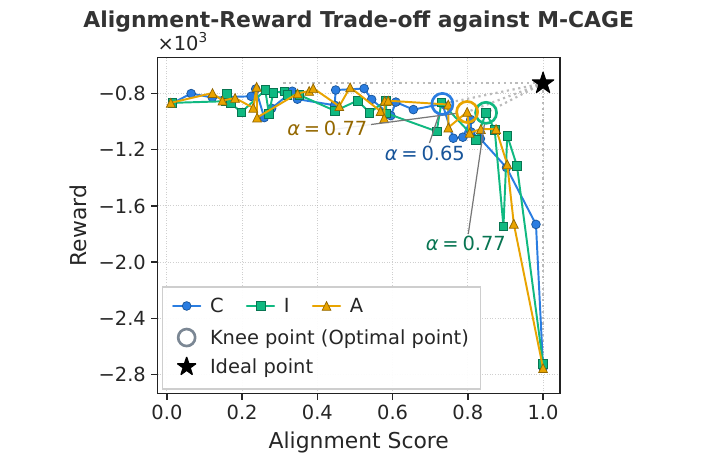}
\includegraphics[width=.4\columnwidth]{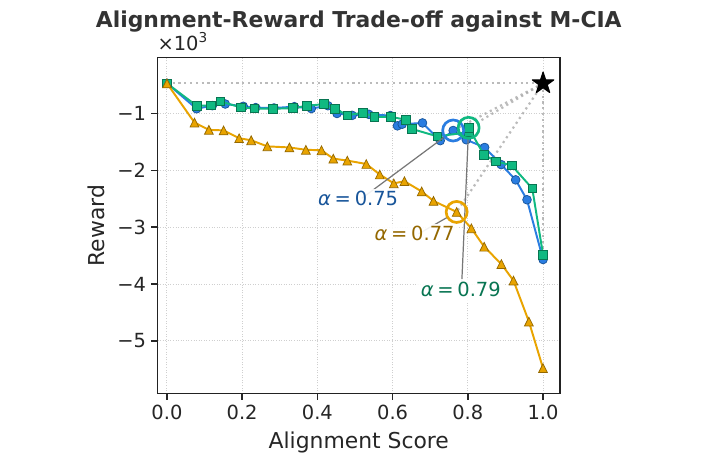}
\vspace{-10pt}
\caption{\protect\rev{Pareto front showing the trade-off between CAGE reward and alignment scores for test-time policy shaping agents on the base scenario. Circles mark knee-points for the CIA attributes ($\alpha^*$). The red M-CIA agent (right) produces a sharply convex trade-off with knee-points at $\alpha \approx 0.75$-$0.79$, while the red M-CAGE agent (left) yields a flatter curve with knee-points at $\alpha \approx 0.65$-$0.77$, reflecting a lower alignment cost when the red agent does not directly target CIA-critical hosts.}}
\label{fig:pareto}
\end{figure}

\rev{The reward cost of fixed-$\alpha$ test-time policy shaping also depends strongly on the adversary's strategy, as illustrated by the Pareto fronts and estimated knee-points in Fig.~\ref{fig:pareto}.} Under the red M-CIA agent, which directly targets CIA-tied hosts, the Pareto front is sharply convex and knee-points occur at moderate $\alpha$ values, beyond which reward degrades steeply. Under the red M-CAGE agent, whose attacks are not CIA-targeted, the Pareto front is substantially flatter, and knee-points shift to higher $\alpha$ values, allowing the defender to sustain strong alignment with modest reward loss. This contrast indicates that the cost of alignment \rev{depends on} the adversary's attack pattern: when the red agent directly targets CIA-tied hosts, the defender must sacrifice more operational reward to maintain the same level of CIA protection. \rev{This higher reward cost motivates activating the alignment-driven policy selectively in response to observed attacks, as in our adaptive approach.}

\rev{This adaptive $\alpha$ improves the reward--alignment trade-off relative to} fixed-$\alpha$ test-time policy shaping at high $\alpha$ values \rev{by avoiding a constant reward cost. Relative to an intermediate fixed setting ($\alpha=0.5$), however, the gain in aggregate alignment score is modest ($\approx$0.52 versus $\approx$0.51 in the base scenario). This modest gain is consistent with the reactive nature of $h(o_t)$, which raises $\alpha_t$ only once a CIA-critical host has already been impacted. \rev{Before this switch, the agent follows the unshaped RL policy. This also} introduces a response delay:} if the red agent compromises a CIA-tied host at step $t$, the blue agent detects this only at step $t+1$, resulting in a brief unprotected window before the policy switches\rev{. The adaptive and fixed-$\alpha$ agents therefore differ in when they prioritize alignment: the adaptive agent increases its alignment weight after detecting impact, whereas the fixed-$\alpha$ agent applies the same weight throughout the episode. These results suggest that improving the policy switching condition in $h(o_t)$ could yield further gains, motivating anticipatory rules that raise $\alpha_t$ on early indicators of compromise rather than on impact alone}.

\del{Fig.~\mbox{\ref{fig:base_CIA_alignment}} (top row) quantifies the test-time alignment-reward trade-off along a Pareto front. Alignment scores for all three CIA attributes scale nearly linearly with $\alpha$ and are balanced across CIA at every network scale tested. While the magnitude of the reward changes with scale (reflecting the increased difficulty in defending larger networks), the reward cost remains convex across scales: moderate $\alpha$ values offer favorable cost-benefit ratios, while $\alpha = 1.0$ guarantees full compliance at a steeper penalty, allowing human operators to tune $\alpha$ at deployment time without retraining.  To find the optimal $\alpha$ value for a given setting, we developed a knee-point estimation technique that can be adapted to each alignment attribute and red agent type.}

\subsection{Generalization to Unseen Adversaries and Network Topologies}\label{sec:results:generalization}

To evaluate whether alignment approaches generalize beyond the base scenario (16 hosts), we train separate PPO agents on three scenarios of increasing network size: $Hosts\_2$ (31 hosts), $Hosts\_3$ (46 hosts), and $Hosts\_4$ (61 hosts), each against the red M-CIA agent (Tab.~\ref{tbl:network_demographics}). Episode duration is scaled proportionally from 200 to 500 steps to accommodate the additional time required by the red agent to traverse larger networks. Furthermore, we also test if alignment approaches generalize to other red agents (i.e., the B-line variants) that were not seen during training.

\begin{figure}[t!]
    \centering
    \includegraphics[width=\columnwidth]{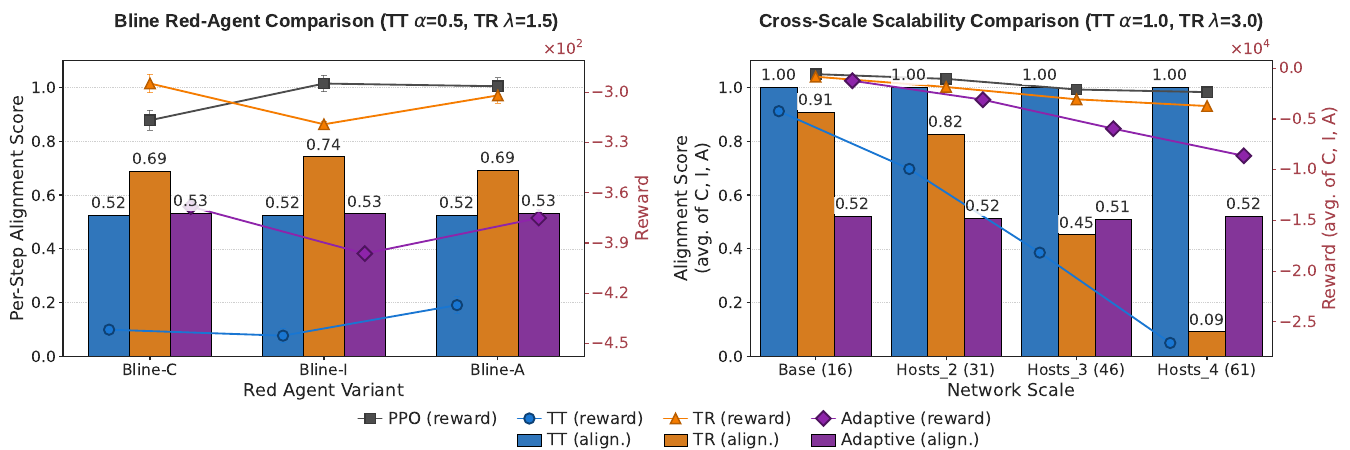}
    \vspace{-10pt}
\caption{Alignment scores (bars, left axis) and CAGE rewards (lines, right axis) for the baseline PPO, test-time policy shaping \proofread{($\alpha=0.5$, $\alpha=1.0$)}, training-time reward shaping \proofread{($\lambda=1.5$, $\lambda=3.0$)}, and adaptive blue agents. (Left) Results \rev{against variants of the red B-line agent}, which target CIA-critical hosts using a direct attack path. CAGE rewards are averaged across the Confidentiality, Integrity, and Availability attributes. (Right) Results \rev{against the red M-CIA agent across four network scales}.}
    \label{fig:scalability_per_network_scale}
\end{figure}

\paragraph{Alignment as network size increases:}
Fig.~\ref{fig:scalability_per_network_scale} compares alignment scores and reward across four network scales (from 16 to 61 hosts). Reward decreases with increasing scale, indicating the challenge of defending larger networks, and the drop is steeper for the adaptive and test-time policy shaping agents. As the standard CAGE reward function does not incentivize protecting CIA-critical hosts, the baseline PPO agent achieves zero alignment scores (not shown).
Agents aligned with training-time reward shaping achieve relatively high alignment scores ($\approx$0.91) in the base scenario; however, we observe alignment rapidly degrade at the largest network scales. In contrast, the test-time policy shaping ($\approx$1.0) and adaptive policy shaping ($\approx$0.51-0.52) agents maintain relatively stable alignment across all four network sizes. We hypothesize that this may be due to gradient collapse from the alignment reward signal during training. \rev{As the network grows, the agent must learn the alignment objective over a larger state and action space, potentially making the alignment reward increasingly sparse or diluted relative to the task reward. In contrast, adaptive policy shaping does not require the agent to learn this trade-off during training, but instead directly increases the influence of the alignment policy when an alignment-relevant impact is observed.} While training-time reward shaping learns effective CIA-aligned policies for the base scenario,
the alignment reward collapses toward zero and the agents revert to near-unaligned behavior at larger network sizes, making the adaptive and test-time policy shaping approaches more effective at these scales. This trend is further illustrated in Fig.~\ref{fig:base_CIA_alignment}, which compares test-time alignment across $\alpha$ (top row) and training-time alignment across $\lambda$ (bottom row) for the base scenario (left column) and the hosts-3 scenario (right column). In the base scenario, increasing either alignment parameter improves the alignment score, with the training-time approach achieving higher task reward than the test-time approach. However, in the larger hosts-3 scenario, the training-time approach fails to learn effective CIA-aligned policies, resulting in substantially lower alignment scores despite maintaining higher task reward. In contrast, the test-time approach continues to improve alignment as $\alpha$ increases, demonstrating greater robustness to increasing network size.

\paragraph{Alignment across targeted red agents:}
To test generalization to unseen adversaries, we evaluate the performance and alignment of blue agents against various B-line red agents, which target CIA-critical hosts using a direct attack path. We note that blue agents were not trained against any of the red B-line variants. Fig.~\ref{fig:scalability_per_network_scale} (left)
shows alignment scores and reward against each B-line variant in the base scenario, using mid-point settings for training-time alignment ($\lambda=1.5$) and test-time alignment ($\alpha=0.5$), to study trade-offs between the two approaches. Across all methods, reward values remain somewhat consistent across red B-line variants. Test-time policy shaping ($\approx$0.51) and adaptive policy shaping ($\approx$0.52) maintain consistent alignment scores across all three variants. In addition,
training-time reward shaping achieves slightly higher alignment scores across B-line red agents ($\approx 0.69$-$0.74$) while maintaining a similar reward to PPO across variants. The baseline PPO agent again achieves zero alignment in all cases, confirming that standard RL training provides no CIA protection under direct CIA-targeted attacks. In terms of reward, test-time policy shaping yields the lowest values among the three approaches. In contrast, the adaptive approach achieves substantially higher reward across novel red adversaries while maintaining similar alignment scores, offering a balanced trade-off without requiring retraining.

%% file: sections/conclusion.tex
\section{Conclusion}
We formulate alignment of autonomous cyber defense agents as test-time policy shaping over a fixed RL policy, \proofread{comparing this approach with} training-time reward shaping 
and a new adaptive policy shaping approach. 
To study this, we focus on alignment to three critical information security attributes or objectives: Confidentiality, Integrity, and Availability.
We show that the \proofread{alignment--reward} trade-off is convex and scale-invariant. 
Our analysis suggests that an ACD agent employing the adaptive alignment approach offers a balanced and practical solution to this trade-off.
Notably, alignment based on training-time reward shaping 
collapses at scale -- 
this result underscores that alignment objectives incorporated during training do not guarantee alignment at deployment. 
Future work includes \proofread{learning functions that map} observations to $\alpha_t$ values for adaptive policy shaping, learning aligned policies from human feedback, and extending our work to multi-attribute alignment with connections to multi-objective reinforcement learning. 

%% file: tables/base.tex
\begin{table*}[b!]
    \centering
    \resizebox{1\textwidth}{!}{%
    \begin{tabular}{l|c|l|c|c|cc|cc|c}
\hline
\multirow{2}{*}{\textbf{Red Agent}} &\multicolumn{2}{|c|}{\raisebox{-6pt}{\textbf{Metric}}} & \multirow{2}{*}{\textbf{BM}} & \multirow{2}{*}{\textbf{PPO}} & \multicolumn{2}{c|}{\textbf{Test-time}} & \multicolumn{2}{c|}{\textbf{Training-time}} & \textbf{Adaptive} \\
& \multicolumn{2}{|c|}{} && & $\alpha=0.5$ & $\alpha=1.0$ & $\lambda=1.5$ & $\lambda=3.0$& $\alpha\in\{0,1\}$  \\
\hline
\multirow{9}{*}{M-CIA} & \multirow{3}{*}{C} & Alignment ($\uparrow$) & 0.00 & 0.00 & 0.53 & \textbf{1.00} & 0.53 & 0.90 & 0.52 \\
 &  & Reward ($\uparrow$) & -3199.12 & \textbf{-587.69} & -991.39 & -3648.42 & -657.89 & -723.16 & -969.98 \\
 &  & Res. Drop ($\downarrow$) & 0.61 & 0.61 & 0.29 & \textbf{0.27} & 0.28 & \textbf{0.27} & 0.28 \\
\cline{2-10}
 & \multirow{3}{*}{I} & Alignment ($\uparrow$) & 0.00 & 0.00 & 0.52 & \textbf{1.00} & 0.53 & 0.94 & 0.52 \\
 &  & Reward ($\uparrow$) & -3199.12 & \textbf{-587.69} & -978.30 & -3575.89 & -635.79 & -653.04 & -903.32 \\
 &  & Res. Drop ($\downarrow$) & 0.61 & 0.61 & 0.53 & \textbf{0.51} & 0.57 & 0.55 & 0.54 \\
\cline{2-10}
 & \multirow{3}{*}{A} & Alignment ($\uparrow$) & 0.00 & 0.00 & 0.53 & \textbf{1.00} & 0.46 & 0.89 & 0.52 \\
 &  & Reward ($\uparrow$) & -3199.12 & \textbf{-587.69} & -1946.82 & -5492.42 & -1007.24 & -1145.55 & -1841.96 \\
 &  & Res. Drop ($\downarrow$) & 0.61 & 0.61 & 0.15 & 0.09 & 0.11 & \textbf{0.01} & 0.08 \\
\hline
\multirow{3}{*}{M-CAGE$^\dagger$} & C & Reward ($\uparrow$) & -3937.68 & -729.19 & -804.28 & -2963.76 & \textbf{-723.64} & -774.40 & -731.28 \\
 & I & Reward ($\uparrow$) & -3937.68 & -729.19 & -810.06 & -2973.65 & -754.40 & -722.87 & \textbf{-716.25} \\
 & A & Reward ($\uparrow$) & -3937.68 & -729.19 & -816.92 & -2974.32 & -928.58 & -955.00 & \textbf{-723.86} \\
\hline
\multirow{3}{*}{Bline-C} & \multirow{3}{*}{C} & Alignment ($\uparrow$) & 0.00 & 0.00 & 0.52 & \textbf{1.00} & 0.69 & 0.97 & 0.53 \\
 &  & Reward ($\uparrow$) & -4039.64 & -316.63 & -441.78 & -3383.29 & \textbf{-294.66} & -375.10 & -368.28 \\
 &  & Res. Drop ($\downarrow$) & 0.85 & 0.83 & 0.28 & 0.16 & 0.08 & \textbf{0.01} & 0.17 \\
\hline
\multirow{3}{*}{Bline-I} & \multirow{3}{*}{I} & Alignment ($\uparrow$) & 0.00 & 0.00 & 0.52 & \textbf{1.00} & 0.74 & 0.97 & 0.53 \\
 &  & Reward ($\uparrow$) & -4039.64 & \textbf{-294.70} & -445.35 & -3368.02 & -319.05 & -304.20 & -396.33 \\
 &  & Res. Drop ($\downarrow$) & 0.68 & 0.66 & 0.22 & 0.13 & 0.06 & \textbf{0.01} & 0.13 \\
\hline
\multirow{3}{*}{Bline-A} & \multirow{3}{*}{A} & Alignment ($\uparrow$) & 0.00 & 0.00 & 0.52 & \textbf{1.00} & 0.69 & 0.98 & 0.53 \\
 &  & Reward ($\uparrow$) & -4039.44 & \textbf{-296.31} & -427.11 & -3339.64 & -301.79 & -346.85 & -375.05 \\
 &  & Res. Drop ($\downarrow$) & 0.65 & 0.63 & 0.21 & 0.12 & 0.07 & \textbf{0.00} & 0.13 \\
\hline

    \end{tabular}%
    }
    \caption{Performance analysis of various blue and red agents on the base scenario (16 hosts, 200 steps, 100 episodes), averaged over five training seeds. Specifically, \textbf{BM} is a Blue Monitor agent that monitors the network only and takes no defensive actions. 
    \textbf{PPO} is the unaligned baseline agent. \textbf{Test-time} is the agent aligned with the test-time policy shaping value $\alpha$. \textbf{Training-time} is the agent aligned with the training-time reward shaping value $\lambda$. \textbf{Adaptive} is the adaptive policy shaping agent. Metrics are quantified with respect to CIA attributes and include \textit{Alignment}: fraction of time steps where the agent chose the C/I/A-optimal action; \textit{Reward}: overall CAGE environment reward;  \textit{Res. Drop}: weighted resilience drop across all CIA attributes. 
    $^\dagger$M-CAGE only targets the operational server (which has no CIA scoring), so only the reward is reported.\del{ See Sec.~\ref{sec:results:baseline} for more details.}}
    \vspace{-10pt}    
    \label{tbl:base}
\end{table*}

%% file: tables/hosts_2.tex
\begin{table*}[h!]
    \centering
    \resizebox{\textwidth}{!}{%
    \begin{tabular}{l|c|l|c|c|cc|cc|c}
\hline
\multirow{2}{*}{\textbf{Red Agent}} &\multicolumn{2}{|c|}{\raisebox{-6pt}{\textbf{Metric}}} & \multirow{2}{*}{\textbf{BM}} & \multirow{2}{*}{\textbf{PPO}} & \multicolumn{2}{c|}{\textbf{Test-time}} & \multicolumn{2}{c|}{\textbf{Training-time}} & \textbf{Adaptive} \\
& \multicolumn{2}{|c|}{} && & $\alpha=0.5$ & $\alpha=1.0$ & $\lambda=1.5$ & $\lambda=3.0$& $\alpha\in\{0,1\}$  \\
\hline
\multirow{9}{*}{M-CIA} & \multirow{3}{*}{C} & Alignment ($\uparrow$) & 0.00 & 0.00 & 0.53 & \textbf{1.00} & 0.31 & 0.72 & 0.52 \\
 &  & Reward ($\uparrow$) & -5483.17 & \textbf{-1053.83} & -2042.26 & -7149.68 & -1336.29 & -1537.66 & -1975.27 \\
 &  & Res. Drop ($\downarrow$) & 0.59 & 0.60 & \textbf{0.30} & \textbf{0.30} & 0.31 & 0.31 & \textbf{0.30} \\
\cline{2-10}
 & \multirow{3}{*}{I} & Alignment ($\uparrow$) & 0.00 & 0.00 & 0.53 & \textbf{1.00} & 0.29 & 0.89 & 0.51 \\
 &  & Reward ($\uparrow$) & -5483.17 & \textbf{-1053.83} & -1973.99 & -7285.76 & -1413.32 & -1508.95 & -1994.00 \\
 &  & Res. Drop ($\downarrow$) & \textbf{0.59} & 0.60 & \textbf{0.59} & \textbf{0.59} & 0.61 & 0.61 & 0.60 \\
\cline{2-10}
 & \multirow{3}{*}{A} & Alignment ($\uparrow$) & 0.00 & 0.00 & 0.52 & \textbf{1.00} & 0.20 & 0.87 & 0.52 \\
 &  & Reward ($\uparrow$) & -5483.17 & \textbf{-1053.83} & -5610.03 & -15365.54 & -1960.21 & -2513.29 & -5353.18 \\
 &  & Res. Drop ($\downarrow$) & 0.59 & 0.60 & 0.14 & 0.08 & 0.23 & \textbf{0.02} & 0.08 \\
\hline
\multirow{3}{*}{M-CAGE$^\dagger$} & C & Reward ($\uparrow$) & -5996.48 & -1406.91 & -1475.90 & -4661.63 & \textbf{-1320.04} & -1333.37 & -1341.90 \\
 & I & Reward ($\uparrow$) & -5996.48 & -1406.91 & -1415.35 & -4654.20 & -1432.73 & -1448.63 & \textbf{-1330.11} \\
 & A & Reward ($\uparrow$) & -5996.48 & -1406.91 & -1448.74 & -4647.87 & -1434.85 & -1916.73 & \textbf{-1342.83} \\
\hline
\multirow{3}{*}{Bline-C} & \multirow{3}{*}{C} & Alignment ($\uparrow$) & 0.00 & 0.00 & 0.52 & \textbf{1.00} & 0.60 & 0.96 & 0.52 \\
 &  & Reward ($\uparrow$) & -5733.15 & \textbf{-503.24} & -653.34 & -5132.07 & -589.24 & -620.76 & -604.73 \\
 &  & Res. Drop ($\downarrow$) & 0.87 & 0.86 & 0.29 & 0.17 & 0.13 & \textbf{0.01} & 0.17 \\
\hline
\multirow{3}{*}{Bline-I} & \multirow{3}{*}{I} & Alignment ($\uparrow$) & 0.00 & 0.00 & 0.52 & \textbf{1.00} & 0.61 & 0.98 & 0.52 \\
 &  & Reward ($\uparrow$) & -5733.15 & \textbf{-507.54} & -652.46 & -5117.10 & -627.78 & -568.27 & -601.41 \\
 &  & Res. Drop ($\downarrow$) & 0.70 & 0.69 & 0.23 & 0.14 & 0.11 & \textbf{0.00} & 0.14 \\
\hline
\multirow{3}{*}{Bline-A} & \multirow{3}{*}{A} & Alignment ($\uparrow$) & 0.00 & 0.00 & 0.52 & \textbf{1.00} & 0.34 & 0.99 & 0.52 \\
 &  & Reward ($\uparrow$) & -5819.76 & -497.10 & -696.17 & -5094.50 & -514.87 & \textbf{-471.68} & -606.04 \\
 &  & Res. Drop ($\downarrow$) & 0.66 & 0.66 & 0.22 & 0.13 & 0.22 & \textbf{0.00} & 0.13 \\
\hline

    \end{tabular}%
    }
    \caption{Performance analysis of various blue and red agents on the Hosts\_2 scenario (31 hosts, 300 steps, 100 episodes), averaged over five training seeds. Specifically, \textbf{BM} is a Blue Monitor agent that monitors the network only and takes no defensive actions. 
    \textbf{PPO} is the unaligned baseline agent. \textbf{Test-time} is the agent aligned with test-time policy shaping value $\alpha$. \textbf{Training-time} is the agent aligned with training-time reward shaping value $\lambda$. \textbf{Adaptive} is the adaptive policy shaping agent. Metrics are quantified with respect to CIA attributes and include \textit{Alignment}: fraction of time steps where the agent chose the C/I/A-optimal action; \textit{Reward}: overall CAGE environment reward; \textit{Res. Drop}: weighted resilience drop across all CIA attributes. 
    $^\dagger$M-CAGE only targets the operational server (which has no CIA scoring), so only reward is reported.}
    \vspace{-10pt}
    \label{tbl:hosts_2}
\end{table*}

%% file: tables/hosts_3.tex
\begin{table*}[b!]
    \centering
    \resizebox{\textwidth}{!}{%
    \begin{tabular}{l|c|l|c|c|cc|cc|c}
\hline
\multirow{2}{*}{\textbf{Red Agent}} &\multicolumn{2}{|c|}{\raisebox{-6pt}{\textbf{Metric}}} & \multirow{2}{*}{\textbf{BM}} & \multirow{2}{*}{\textbf{PPO}} & \multicolumn{2}{c|}{\textbf{Test-time}} & \multicolumn{2}{c|}{\textbf{Training-time}} & \textbf{Adaptive} \\
& \multicolumn{2}{|c|}{} && & $\alpha=0.5$ & $\alpha=1.0$ & $\lambda=1.5$ & $\lambda=3.0$& $\alpha\in\{0,1\}$  \\
\hline
\multirow{9}{*}{M-CIA} & \multirow{3}{*}{C} & Alignment ($\uparrow$) & 0.00 & 0.00 & 0.51 & \textbf{1.00} & 0.05 & 0.55 & 0.51 \\
 &  & Reward ($\uparrow$) & -8343.21 & \textbf{-2102.34} & -3655.44 & -11986.94 & -2762.10 & -3135.83 & -3264.31 \\
 &  & Res. Drop ($\downarrow$) & 0.56 & 0.59 & 0.31 & \textbf{0.30} & 0.36 & 0.32 & 0.31 \\
\cline{2-10}
 & \multirow{3}{*}{I} & Alignment ($\uparrow$) & 0.00 & 0.00 & 0.51 & \textbf{1.00} & 0.04 & 0.74 & 0.51 \\
 &  & Reward ($\uparrow$) & -8343.21 & \textbf{-2102.34} & -3460.37 & -12834.76 & -2658.48 & -3093.05 & -3413.15 \\
 &  & Res. Drop ($\downarrow$) & \textbf{0.56} & 0.59 & 0.61 & 0.59 & 0.63 & 0.62 & 0.62 \\
\cline{2-10}
 & \multirow{3}{*}{A} & Alignment ($\uparrow$) & 0.00 & 0.00 & 0.50 & \textbf{1.00} & 0.01 & 0.06 & 0.51 \\
 &  & Reward ($\uparrow$) & -8343.21 & -2102.34 & -11627.07 & -29725.86 & \textbf{-2052.61} & -2974.49 & -11206.41 \\
 &  & Res. Drop ($\downarrow$) & 0.56 & 0.59 & 0.14 & \textbf{0.08} & 0.58 & 0.48 & \textbf{0.08} \\
\hline
\multirow{3}{*}{M-CAGE$^\dagger$} & C & Reward ($\uparrow$) & -8510.52 & -2335.89 & -2360.42 & -6943.90 & \textbf{-2217.40} & -2303.05 & -2284.57 \\
 & I & Reward ($\uparrow$) & -8510.52 & -2335.89 & -2380.96 & -6925.75 & \textbf{-2208.97} & -2442.47 & -2278.56 \\
 & A & Reward ($\uparrow$) & -8510.52 & -2335.89 & -2385.60 & -6894.42 & \textbf{-2189.82} & -2528.40 & -2294.10 \\
\hline
\multirow{3}{*}{Bline-C} & \multirow{3}{*}{C} & Alignment ($\uparrow$) & 0.00 & 0.00 & 0.52 & \textbf{1.00} & 0.25 & 0.92 & 0.52 \\
 &  & Reward ($\uparrow$) & -8444.50 & \textbf{-1033.15} & -1034.86 & -7381.22 & -1202.52 & -1254.60 & -1314.10 \\
 &  & Res. Drop ($\downarrow$) & 0.88 & 0.87 & 0.29 & 0.18 & 0.39 & \textbf{0.02} & 0.18 \\
\hline
\multirow{3}{*}{Bline-I} & \multirow{3}{*}{I} & Alignment ($\uparrow$) & 0.00 & 0.00 & 0.52 & \textbf{1.00} & 0.32 & 0.91 & 0.52 \\
 &  & Reward ($\uparrow$) & -8444.50 & \textbf{-1103.63} & -1233.28 & -7378.89 & -1174.75 & -1204.66 & -1314.53 \\
 &  & Res. Drop ($\downarrow$) & 0.70 & 0.70 & 0.23 & 0.14 & 0.28 & \textbf{0.02} & 0.14 \\
\hline
\multirow{3}{*}{Bline-A} & \multirow{3}{*}{A} & Alignment ($\uparrow$) & 0.00 & 0.00 & 0.52 & \textbf{1.00} & 0.05 & 0.10 & 0.52 \\
 &  & Reward ($\uparrow$) & -8483.08 & \textbf{-1062.13} & -1115.72 & -7312.82 & -1106.50 & -1272.37 & -1304.62 \\
 &  & Res. Drop ($\downarrow$) & 0.69 & 0.68 & 0.23 & 0.14 & 0.56 & 0.53 & \textbf{0.13} \\
\hline

    \end{tabular}%
    }
    \caption{Performance analysis of various blue and red agents on the Hosts\_3 scenario (46 hosts, 400 steps, 100 episodes), averaged over five training seeds. Specifically, \textbf{BM} is a Blue Monitor agent that monitors the network only and takes no defensive actions. 
    \textbf{PPO} is the unaligned baseline agent. \textbf{Test-time} is the agent aligned with test-time policy shaping value $\alpha$. \textbf{Training-time} is the agent aligned with training-time reward shaping value $\lambda$. \textbf{Adaptive} is the adaptive policy shaping agent. Metrics are quantified with respect to CIA attributes and include \textit{Alignment}: fraction of time steps where the agent chose the C/I/A-optimal action; \textit{Reward}: overall CAGE environment reward; \textit{Res. Drop}: weighted resilience drop across all CIA attributes. 
    $^\dagger$M-CAGE only targets the operational server (which has no CIA scoring), so only reward is reported.}
    \vspace{-10pt}
    \label{tbl:hosts_3}
\end{table*}

%% file: tables/hosts_4.tex
\begin{table*}[b!]
    \centering
    \resizebox{\textwidth}{!}{%
    \begin{tabular}{l|c|l|c|c|cc|cc|c}
\hline
\multirow{2}{*}{\textbf{Red Agent}} &\multicolumn{2}{|c|}{\raisebox{-6pt}{\textbf{Metric}}} & \multirow{2}{*}{\textbf{BM}} & \multirow{2}{*}{\textbf{PPO}} & \multicolumn{2}{c|}{\textbf{Test-time}} & \multicolumn{2}{c|}{\textbf{Training-time}} & \textbf{Adaptive} \\
& \multicolumn{2}{|c|}{} && & $\alpha=0.5$ & $\alpha=1.0$ & $\lambda=1.5$ & $\lambda=3.0$& $\alpha\in\{0,1\}$  \\
\hline
\multirow{9}{*}{M-CIA} & \multirow{3}{*}{C} & Alignment ($\uparrow$) & 0.00 & 0.00 & 0.50 & \textbf{1.00} & 0.00 & 0.06 & 0.52 \\
 &  & Reward ($\uparrow$) & -9247.90 & \textbf{-2358.14} & -4559.83 & -16991.56 & -2852.47 & -4028.11 & -4332.24 \\
 &  & Res. Drop ($\downarrow$) & 0.63 & 0.61 & \textbf{0.31} & \textbf{0.31} & 0.54 & 0.35 & 0.32 \\
\cline{2-10}
 & \multirow{3}{*}{I} & Alignment ($\uparrow$) & 0.00 & 0.00 & 0.51 & \textbf{1.00} & 0.01 & 0.21 & 0.52 \\
 &  & Reward ($\uparrow$) & -9247.90 & \textbf{-2358.14} & -4638.58 & -16487.74 & -2973.00 & -4224.32 & -4499.55 \\
 &  & Res. Drop ($\downarrow$) & 0.63 & 0.61 & 0.61 & 0.61 & \textbf{0.60} & 0.63 & 0.62 \\
\cline{2-10}
 & \multirow{3}{*}{A} & Alignment ($\uparrow$) & 0.00 & 0.00 & 0.51 & \textbf{1.00} & 0.00 & 0.00 & 0.52 \\
 &  & Reward ($\uparrow$) & -9247.90 & -2358.14 & -17833.46 & -47764.26 & \textbf{-2173.33} & -2938.43 & -17073.98 \\
 &  & Res. Drop ($\downarrow$) & 0.63 & 0.61 & 0.14 & \textbf{0.08} & 0.61 & 0.61 & \textbf{0.08} \\
\hline
\multirow{3}{*}{M-CAGE$^\dagger$} & C & Reward ($\uparrow$) & -10253.23 & -2910.22 & -2694.31 & -8149.23 & -2952.69 & -2783.82 & \textbf{-2629.99} \\
 & I & Reward ($\uparrow$) & -10253.23 & -2910.22 & -2721.87 & -8115.50 & -2804.95 & -2853.31 & \textbf{-2663.83} \\
 & A & Reward ($\uparrow$) & -10253.23 & -2910.22 & -2730.70 & -8145.74 & \textbf{-2449.88} & -3320.91 & -2610.66 \\
\hline
\multirow{3}{*}{Bline-C} & \multirow{3}{*}{C} & Alignment ($\uparrow$) & 0.00 & 0.00 & 0.51 & \textbf{1.00} & 0.02 & 0.27 & 0.52 \\
 &  & Reward ($\uparrow$) & -9506.02 & -1083.17 & -1325.20 & -8442.14 & -1202.86 & \textbf{-1045.29} & -1318.83 \\
 &  & Res. Drop ($\downarrow$) & 0.88 & 0.88 & 0.29 & \textbf{0.18} & 0.82 & 0.38 & \textbf{0.18} \\
\hline
\multirow{3}{*}{Bline-I} & \multirow{3}{*}{I} & Alignment ($\uparrow$) & 0.00 & 0.00 & 0.51 & \textbf{1.00} & 0.06 & 0.54 & 0.52 \\
 &  & Reward ($\uparrow$) & -9506.02 & \textbf{-1106.57} & -1305.74 & -8555.80 & -1141.21 & -1267.65 & -1279.65 \\
 &  & Res. Drop ($\downarrow$) & 0.70 & 0.70 & 0.23 & \textbf{0.14} & 0.54 & \textbf{0.14} & \textbf{0.14} \\
\hline
\multirow{3}{*}{Bline-A} & \multirow{3}{*}{A} & Alignment ($\uparrow$) & 0.00 & 0.00 & 0.51 & \textbf{1.00} & 0.00 & 0.00 & 0.52 \\
 &  & Reward ($\uparrow$) & -9408.34 & \textbf{-1014.99} & -1593.43 & -8594.74 & -1252.23 & -1141.46 & -1318.22 \\
 &  & Res. Drop ($\downarrow$) & 0.69 & 0.69 & 0.23 & \textbf{0.14} & 0.67 & 0.67 & \textbf{0.14} \\
\hline

    \end{tabular}%
    }
    \vspace{-5pt}
    \caption{Performance analysis of blue and red agents on the Hosts\_4 scenario (61 hosts, 500 steps, 100 episodes), averaged over five training seeds. Specifically, \textbf{BM} is a Blue Monitor agent that monitors the network only and takes no defensive actions. 
    \textbf{PPO} is the unaligned baseline agent. \textbf{Test-time} is the agent aligned with test-time policy shaping value $\alpha$. \textbf{Training-time} is the agent aligned with training-time reward shaping value $\lambda$. \textbf{Adaptive} is the adaptive policy shaping agent. Metrics are quantified with respect to CIA attributes and include \textit{Alignment}: fraction of time steps where the agent chose the C/I/A-optimal action; \textit{Reward}: overall CAGE environment reward; \textit{Res. Drop}: weighted resilience drop across all CIA attributes. 
    $^\dagger$M-CAGE only targets the operational server (which has no CIA scoring), so only reward is reported.}
    \vspace{-10pt}
    \label{tbl:hosts_4}
\end{table*}